\documentclass[5p,times]{elsarticle}
\journal{European Journal of Control}

\usepackage{amsthm}
\newtheoremstyle{mystyle}
  {}
  {}
  {}
  {}
  {\bfseries}
  {.}
  { }
  {}

\theoremstyle{mystyle}

\usepackage{amsmath} 
\usepackage{amssymb}  
\usepackage{blindtext}

\usepackage{etoolbox}
\usepackage{lipsum} 

\usepackage{siunitx}
\usepackage{pifont}
\usepackage{graphicx}
\usepackage{amsmath, textcomp}
\usepackage{algorithm} 
\usepackage{cite}
\usepackage{hyperref}
\usepackage{booktabs}
\usepackage{tabularx}
\usepackage{cleveref}

\usepackage{url}
\usepackage{subcaption} 
\usepackage{algpseudocode}
\usepackage{paralist}
\usepackage{color}
\usepackage{relsize}
\usepackage{stmaryrd}
\usepackage{epsfig} 
\usepackage{bm} 
\usepackage{listings,lstautogobble}
\usepackage{color}
\usepackage{mathrsfs}
\usepackage{xspace}
\usepackage{float}
\usepackage{verbatim}
\usepackage{epstopdf}
\usepackage{listings}
\usepackage{parcolumns}
\usepackage{color}
\usepackage{mathtools}
\usepackage{mathrsfs}
\usepackage{xspace}
\usepackage{soul}
\usepackage{float}
\usepackage{verbatim}
\usepackage[utf8]{inputenc}
\usepackage{calrsfs}
\usepackage{rotating,multirow}
\usepackage{caption}

\usepackage{amsfonts}
\usepackage[most]{tcolorbox}
\usepackage{float}
\usepackage[T1]{fontenc}

\allowdisplaybreaks

\usepackage{todonotes}
\usepackage{cleveref}
\usepackage{tikz}
\usetikzlibrary{external}
\usepackage{pgfplots}
\usetikzlibrary{matrix}
\usepgfplotslibrary{groupplots}
\pgfplotsset{compat=newest}
\usepackage{pgfplotstable}

\DeclareMathAlphabet{\mathcal}{OMS}{cmsy}{m}{n}

\DeclarePairedDelimiter{\abs}{\lvert}{\rvert}

\usepackage{enumitem}

\newcommand{\R}{{\mathbb{R}}}

\newtheorem{theorem}{\textbf{Theorem}}
\newtheorem{lemma}{Lemma}

\newtheorem{problem}{Problem}

\newtheorem{definition}{Definition}
\newtheorem{example}{Example}

\newtheorem{assumption}{Assumption}
\newtheorem{remark}{\textbf{Remark}}

\newcommand{\enc}[1]{\llbracket #1 \rrbracket}

\newcommand{\norm}[1]{\left\Vert#1\right\Vert}
\renewcommand{\matrix}[1]{\begin{bmatrix}#1\end{bmatrix}}

\usepackage{xcolor}

\newtcblisting[auto counter]{samplelisting}[2][]{sharp corners, 
    fonttitle=\bfseries, colframe=gray, listing only,
    listing options={basicstyle=\fontsize{7.3}{12}\ttfamily,xleftmargin=0em, aboveskip=0em, belowcaptionskip=0em,,
    belowskip=0em,showstringspaces=false,
    breaklines=false,language=java},   
    title=Function \thetcbcounter: #2, #1}

\usepackage{eso-pic}
\newcommand{\zono}[1]{\langle#1\rangle}
\newcommand{\setdef}[2][]{
	\bigg\{
	\ifblank{#1}{}{#1 \hspace{.1cm} \Big| \hspace{.1cm}}
	#2
	\bigg\}
}

\newcommand{\f}[2]{#1\left(#2\right)}
\newcommand{\set}[1]{\mathcal{#1}}
\begin{document}

\begin{frontmatter}
\title{Privacy-Preserving Set-Based Estimation Using Partially Homomorphic Encryption}



\author[add1]{Amr Alanwar\corref{cor1}}
\ead{aalanwar@constructor.university}
\author[add2]{Victor Gaßmann}
\ead{victor.gassmann@tum.de}
\author[add3]{Xingkang He}
\ead{xingkang@kth.se}
\author[add4]{Hazem Said}
\ead{hazem.said@eng.asu.edu.eg}
\author[add3]{Henrik Sandberg}
\ead{hsan@kth.se}
\author[add3]{Karl Henrik Johansson}
\ead{kallej@kth.se}
\author[add2]{Matthias Althoff}
\ead{althoff@tum.de}

\address[add1]{School of Computer Science and Engineering, Constructor University, Germany}
\address[add2]{Department of Computer Engineering, Technical University of Munich, Germany}
\address[add3]{Division of Decision and Control Systems, KTH Royal Institute of Technology, Sweden}
\address[add4]{Department of Computer Engineering, Ain Shams University, Egypt}

\cortext[cor1]{Corresponding author.}

\begin{abstract}

The set-based estimation has gained a lot of attention due to its ability to guarantee state enclosures for safety-critical systems. However, collecting measurements from distributed sensors often requires outsourcing the set-based operations to an aggregator node, raising many privacy concerns. To address this problem, we present set-based estimation protocols using partially homomorphic encryption that preserve the privacy of the measurements and sets bounding the estimates. We consider a linear discrete-time dynamical system with bounded modeling and measurement uncertainties. Sets are represented by zonotopes and constrained zonotopes as they can compactly represent high-dimensional sets and are closed under linear maps and Minkowski addition. By selectively encrypting parameters of the set representations, we establish the notion of encrypted sets and intersect sets in the encrypted domain, which enables guaranteed state estimation while ensuring privacy. In particular, we show that our protocols achieve computational privacy using the cryptographic notion of computational indistinguishability. We demonstrate the efficiency of our approach by localizing a real mobile quadcopter using ultra-wideband wireless devices. 



\end{abstract}

\begin{keyword}
Set-based estimation, homomorphic encryption, zonotopes, constrained zonotopes. 
\end{keyword}
\end{frontmatter}



\section{Introduction}
State estimation from noisy measurements is of great importance in many areas, such as navigation, communication, and remote sensing. Many of these applications are based on prior knowledge of noise distributions. However, assumed noise distributions are not always sufficiently accurate or even unknown. 
Furthermore, safety-critical applications require guaranteed state inclusion in a bounded set to provably avoid unsafe sets. This motivates the need for set-based estimation, which estimates the set of all possible system states when input disturbances and observation errors are unknown but belong to given bounded sets \citep{conf:setmem1971}. 
%
Set-based estimators are used in many applications, such as underwater robotics \citep{conf:setmem2water}, fault detection \citep{conf:set_fault1,conf:faultcombastel1}, leader-follower problems \citep{conf:setmem3eventleader}, and localization \citep{conf:setloc}. We refer the reader to \citep{conf:AlthoffJagatjournal} and references therein for more related work on set-based estimation. 





%


Some state estimation algorithms require measurements ma-de by a set of spatially distributed sensors. For instance, cellular signals from distributed mobile devices can be measured by base stations to estimate the targeted device location \citep{conf:proloc}. Situational awareness in safe autonomous driving requires collecting measurements from distributed vehicles and infrastructure nodes \citep{conf:vandana}. 
These computations require cloud-based services that aggregate and process gathered information to provide estimates with guarantees. 
However, this often requires that clients disclose sensitive information to the cloud to receive appropriate control decisions. 
This causes security vulnerabilities \citep{conf:attacksautomotive,conf:vinyl}, especially when sensors do not belong to the same trust zone in which members of the same organization trust each other. 
For this reason, we focus on set-based estimation in the cloud with estimation and privacy guarantees.  

\subsection{Related Work}\label{sec:related}

There exist three types of set-based observers: strip-based observers,
set-propagation observers, and interval observers \citep{conf:AlthoffJagatjournal}. Since we will use strip-based observers in this work, we focus our literature review on this observer type and refer the interested reader to \citep{conf:AlthoffJagatjournal} for the other observer types. Strip-based observers intersect the propagated set of states with the set of states consistent with the next measurement to obtain the next set of possible states. 
%
%
%
%
The set representation is essential to obtain a good computational complexity ratio and the estimated sets' achieved tightness. 
Ellipsoids are explored in \citep{conf:setmem1971,conf:Ellipsoidalstate,conf:set-ellipsoida}, where the computations are generally efficient but not exact for the Minkowski sums. A new geometric method based on the Minkowski sum is proposed in \citep{conf:ellipsoidaldistributed} to produce a distributed ellipsoidal estimation. Zonotopes \citep{conf:zono1998} are a special class of polytopes for which one can efficiently compute linear maps and Minkowski sums -- both are important operations for set-based observers. Set-membership using zonotopes is explored in~\citep{garcia2020guaranteed}. A novel zonotope intersection method and a new distributed set-based estimator were proposed in \citep{conf:dis-diff}. A distributed zonotopic and Gaussian Kalman filter is proposed in \citep{conf:symboliczono}, where each network node implements a local state estimator using zonotopes and Gaussian noise mergers. Polytopes~\citep{blesa2012robust} and orthotopes~\citep{belforte1990parameter,hamdi2016orthotopic} were explored as well.

Related work on set-based estimation does not provide privacy guarantees. Homomorphic encryption allows processing over encrypted data and has been used as a countermeasure for cloud-side information leakage, enabling useful tasks to be accomplished while keeping the data confidential from untrusted parties. 
Over the past few years, a significant effort in the form of a homomorphic library~\citep{gentry:homolib} has been made to make fully homomorphic encryption practical. Homomorphic encryption has been used for computationally expensive tasks over genome data~\citep{kristin:homo} and classification over encrypted data~\citep{bost:homo}. However, fully homomorphic encryption remains impractical for real-time estimation \citep[Section 2.10.1]{conf:locthes}. That said, partial homomorphic encryption methods are more promising and have been used for encrypted control \citep{conf:alex-control,conf:multhomo}, image processing \citep{conf:crypt}, estimation \citep{conf:proloc,conf:for_multi_estimation}, deep learning \citep{conf:for_deeplearning}, optimization \citep{conf:gradhomo}, and ride-sharing \citep{conf:ridesharing}. 
%
%

A related technique to our work is differential privacy~\citep{conf:distributeddifferentialpr,conf:for_diffprivacy}, which relies on the addition of structured noise to the data before sharing it, which preserves privacy. Variants of this sche-me, such as local differential privacy~\citep{duchi:localdp2,conf:disoptdiffpr} and geo-indisting-uishability \citep{bordenabe:geo}, have been designed to ensure differential privacy for location data. However, the privacy guarantees of these methods are often achieved at the expense of accuracy \citep{conf:diffprvcost}. In other words, the added structured noise results in a loss of estimation accuracy, making it unsuitable for use in safety-critical systems. To overcome the addition of excessive noise, a combination of homomorphic encryption with distributed noise has been proposed in~\citep{shi:privacy}, where each estimator generated its share of the aggregated noise required for differential privacy~\citep{dwork:noise} and sent encrypted and obfuscated data to the aggregator.

\subsection{Contributions}

To the best of our knowledge, for the first time, we leverage a partially homomorphic cryptosystem to calculate encrypted sets that enclose states based on encrypted measurements and estimates from sensors or sensor groups. This work introduces two protocols providing state inclusion and privacy guarantees. 
In particular, we show that our protocols achieve computational privacy using computational indistinguishability against different coalitions of participated entities. We leverage state-of-art state estimation techniques in combination with homomorphic encryption to provide privacy-preserving set-based estimation protocols with security guarantees. Our entire code and data are available online\footnote{\href{https://github.com/aalanwar/Encrypted-set-based-estimation}{https://github.com/aalanwar/Encrypted-set-based-estimation}}. 

More specifically, we make the following contributions:

\begin{figure*}[t]
 \begin{subfigure}{.33\textwidth}
  \centering
  \includegraphics[width=.4\linewidth]{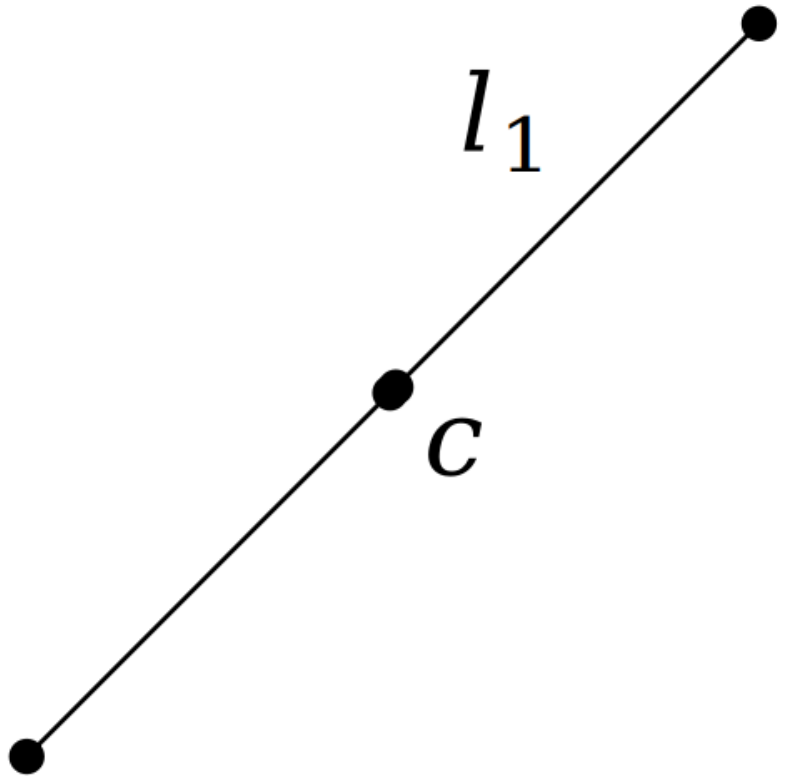}
  \caption{$c \boxplus l_1$}
  \label{fig:zono1}
\end{subfigure}%
\begin{subfigure}{.33\textwidth}
  \centering
  \includegraphics[width=.4\linewidth]{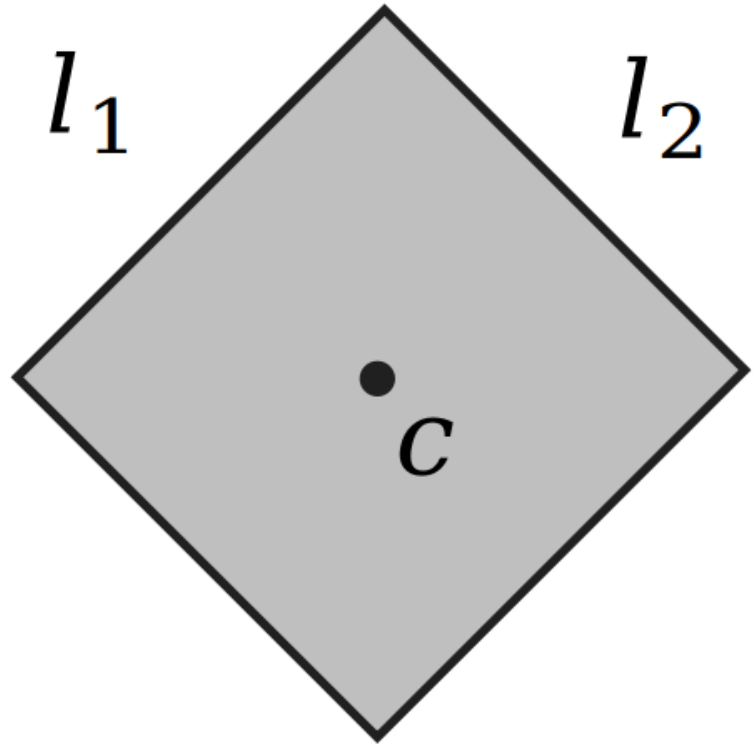}
  \caption{$c \boxplus l_1 \boxplus l_2$ }
  \label{fig:zono2}
\end{subfigure}
\begin{subfigure}{.33\textwidth}
  \centering
  \includegraphics[width=.45\linewidth]{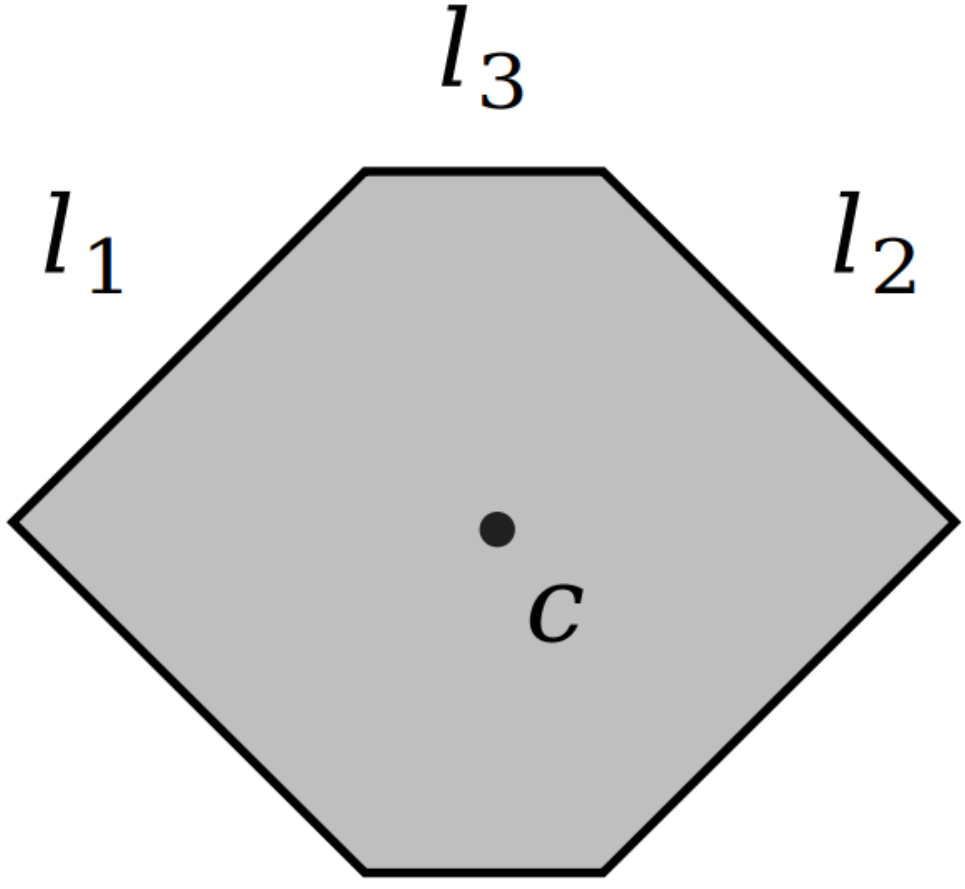}
  \caption{$c \boxplus l_1 \boxplus l_2 \boxplus l_3$}
  \label{fig:zono3}
\end{subfigure}%
\caption{Construction of a zonotope.}
\label{fig:zonocont}
\vspace{-3mm}
\end{figure*}

\begin{itemize}
    \item We encrypt a set of states using a partially homomorphic cryptosystem with different levels of privacy based on selective encryption and geometric features of the chosen set representation.  
    \item We present two set-based estimation protocols which preserve privacy between sensor and sensor groups.  
    \item We prove security guarantees of the two protocols against different coalitions, using formal cryptographic definitions of computational indistinguishability for protecting the estimated set position (Theorems~\ref{th:alg1pricol} and \ref{th:alg2pricol}) and protecting the estimated set position and shape (Theorems~\ref{th:alg3pricol} and \ref{th:alg4pricol}).
\end{itemize}

\subsection{Outline}
The paper is organized as follows: In Section~\ref{sec:prel}, we provide the necessary preliminaries. We formulate the problem and set our privacy goals in Section~\ref{sec:define}. After proposing the notion of encrypted sets in Section~\ref{sec:propsolu}, we introduce protocols to privately bound the state among distributed sensors in Section~\ref{sec:alg1} and then among sensor groups in Section~\ref{sec:alg2}. Finally, we evaluate the proposed protocols in Section~\ref{sec:eval} and conclude this paper with Section~\ref{sec:con}. 

\subsection{Notation}
Vectors and scalars are denoted by lowercase letters, matrices are denoted by uppercase letters, the real and natural numbers are denoted by $\mathbb{R}$ and $\mathbb{N}$. We denote the set of positive real and positive natural numbers by $\mathbb{R}^+$ and $\mathbb{N}^+$, respectively, and all other continuous sets are denoted by calligraphic letters. For a given matrix $M\in\mathbb{R}^{o\times k}$, its Frobenius norm is given by $\norm{M}_F = \sqrt{\f{\text{tr}}{M^TM}}$. For two sets $\set{M}_1\subseteq \mathbb{R}^q$, and $\set{M}_2\subseteq \mathbb{R}^q$, the Minkowski sum and the intersection are denoted by $\set{M}_1 \boxplus \set{M}_2$ and $\set{M}_1 \cap  \set{M}_2$, respectively. For a set $\set{M}\subseteq\mathbb{R}^q$, its linear map is denoted by $L\set{M}$, where $L\in\mathbb{R}^{v\times q}$. For a given matrix $M$ (can also be a vector or scalar), we denote with $\enc{M}$ the encrypted value of $M$. For given vectors $a_1$ and $a_2$ of same dimension, we denote with $\enc{a_1}\oplus \enc{a_2}$ and $\enc{a_1}\ominus \enc{a_2}$ the sum and difference over the encrypted values of $a_1$ and $a_2$, respectively. For two real scalars $a$ and $b$, we denote with $a\otimes \enc{b}$ the multiplication of the encrypted scalar $b$ with the unencrypted scalar $a$. We denote probability of an event $E$ by $\text{Pr}[E]$. The cardinality of a set $\mathcal{M}$ is denoted by $\abs{\mathcal{M}}$. For a given vector $x\in\mathbb{R}^p$, the $i$-th component of $x$ is denoted by $x_{[i]}\in\mathbb{R}$. We denote the reduce operator returning an over-approximative zonotope with $q$ generators by $\downarrow_q$.





\section{Preliminaries} \label{sec:prel}

In this section, we review the required preliminaries.

\subsection{Set Representations and Set-Based Estimation}\label{sec:setbased}

We define the following set representations:


\begin{definition}
\textbf{(Zonotope)} (\citep{conf:zono1998}) An $n$-dimensional zonotope $\mathcal{Z}$ is defined as 
\begin{equation}
\mathcal{Z} = \setdef[x\in\mathbb{R}^n]{x=c+G\beta, \ \norm{\beta}_\infty\leq 1},
    \label{equ:zonoDef}
\end{equation}
%
where $c \in \R^n$ is the center, $G$ $\in$ $\R^{n \times e}$ is the generator matrix of the zonotope, and $\beta\in\mathbb{R}^e$ is the vector of zonotope factors. For later use, we represent the zonotope by $\mathcal{Z} = \zono{c,G}$. Note that $G$ consists of $e$ generators $g^{(i)} \in \R^n$, $i=1,..,e$, such that $G=[g^{(1)},...,g^{(e)}]$.
\label{df:zonotope}
\end{definition}


This definition can be interpreted as the Minkowski sum of a finite set of line segments $l_i = \setdef[g^{(i)}\beta_i\in\mathbb{R}^n]{\abs{\beta_i}\leq 1}$, where $i\in\{1,...,e\}$ (see Figure~\ref{fig:zonocont}). For a zonotope $\mathcal{Z}=\zono{c,G}$, we denote its F-radius as $\norm{G}_F$. Given two zonotopes $\mathcal{Z}_1=\zono{c_1,G_1}$ and $\mathcal{Z}_2=\zono{c_2,G_2}$, the Minkowski sum is computed as $\mathcal{Z}_1 \boxplus \mathcal{Z}_2=\zono{c_1 + c_2, [G_1, G_2]}$, whereas the linear map is computed as $L \mathcal{Z}_1 = \zono{L c_1, L G_1}$ \citep{conf:althoffthesis}.
    
 

\begin{definition}\label{df:contzono}
\textbf{(Constrained Zonotope)} (\citep[Prop. 1]{conf:const_zono}) An $n$-dimensional constrained zonotope is defined as
\begin{equation}\label{eq:conszono}
    \mathcal{C} = \setdef[x\in\mathbb{R}^n]{x=c+G\beta, \ A\beta=b, \, \norm{\beta}_\infty\leq 1}, 
\end{equation}
where $c \in \R^n$ is the center, $G$ $\in$ $\R^{n \times n_g}$ is the generator matrix, $\beta \in \mathbb{R}^{n_g}$, and $A \in $ $\R^{n_c \times n_g}$ and $b \in \R^{n_c}$ constitute the constraints. In short, we write $\mathcal{C}= \zono{c,G,A,b}$.
\end{definition}

\begin{figure*}[h]
    \centering
    \includegraphics[scale=0.9]{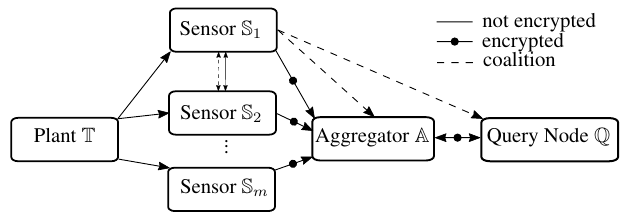}
    \caption{Diagram for the considered setup in Problem \ref{pb:sen}.}
	\label{fig:setup}
	\vspace{-1mm}
\end{figure*}

\begin{figure*}[h]
    \centering
    \includegraphics[scale=0.9]{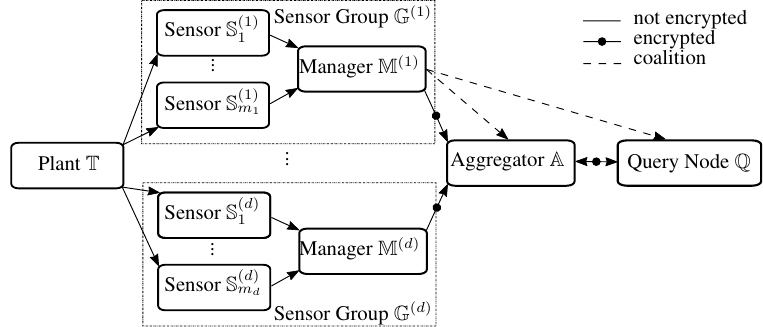}
    \caption{Diagram for the considered setup in Problem \ref{pb:sengrp}.}
	\label{fig:setupGrp}
	\vspace{-4mm}
\end{figure*}

\subsection{Paillier Homomorphic Cryptosystem and Privacy Definitions} \label{sec:PAcrypto}

A homomorphic cryptosystem supports computation over encrypted data.
Our protocols heavily rely on Paillier additive homomorphic cryptosystems~\citep{phe:paillier}, which is a probabilistic public key cryptography scheme. The Paillier cryptosystem supports
\begin{align}
&\textsc{Decrypt}_{\text{sk}}(\enc{a} \oplus \enc{b})  = a + b, \label{eq:pailadd}\\
& \textsc{Decrypt}_{\text{sk}}(a \otimes \enc{b}) = a \cdot b, \label{eq:pailmul}
\end{align}
where $\textit{sk}$ is the private key associated with the public key $\textit{pk}$ used for encryption. 
We will omit the symbol $\otimes$ when the type of multiplication can be inferred from the context. Our proposed protocols can utilize different homomorphic encryption schemes instead of the Paillier cryptosystem as long as it supports the same functionality.

Homomorphic encryption does not support float numbers. The naive solution is multiplying the float number by $10^f$ where $f$ is the number of floating digits \citep{conf:cloutdquadratic,conf:intermittentStateReset,conf:convert2integer}. However, the recursive execution of the estimator or the controller generally requires recursive multiplication with fractional numbers. This approach requires truncating the significance of the state from time to time to avoid overflow. Such truncation might lead to computation errors and fast overflow. We can not use a solution that introduces computation errors because we provide safety and set containment guarantees. To overcome this limitation, we represent float numbers by a positive integer exponent and an integer mantissa, as we did in our previous work \citep{conf:crypt}. This representation provides exact computations. However, it still suffers from overflows after some iterations.


We define $\{0,1\}^\star$ as a sequence of bits of unspecified length. An ensemble $X =\{X_o\}_{o \in \mathbb{N}}$  is a sequence of random variables $X_o$ ranging over strings of  bits of polynomial length in $o$. We need the following definitions in our privacy proofs. 

\begin{definition}
\label{def:comindis}
\textbf{(Computationally Indistinguishable)} (\citep[p.105]{conf:cryptoref})
The ensembles $X = \{X_o\}_{o \in \mathbb{N}}$ and $Y = \{Y_o\}_{o \in \mathbb{N}}$ are computationally indistinguishable, denoted $X\stackrel{c}{\equiv}Y$, if for every probabilistic polynomial-time algorithm D, every positive polynomial $p:\mathbb{N}^+\mapsto \mathbb{R}^+$, and all sufficiently large $o$, it holds that 
\begin{equation}
    \Big| \text{Pr}[D(X_o)=1] - \text{Pr}[D(Y_o)=1] \Big| < \frac{1}{p(o)}.
\end{equation}
\end{definition}

In other words, given an algorithm $D$, we consider the probability that $D$ outputs 1 given an ensemble taken from the two random variables $X_o$ and $Y_o$ as input. Then,  we say $X\stackrel{c}{\equiv}Y$ if no efficient algorithm can tell the difference between them except with small probability $\frac{1}{p(o)}$. 

\begin{definition} 
\label{def:ExecView}
\textbf{(Execution View)}
Let $f:\mathbb{R}^o\mapsto \mathbb{R}^o$ be a deterministic polynomial-time function and $\Pi$ a multi-party protocol computing $f(\bar{x})$, where $\bar{x} \in \mathbb{R}^o$. The view of the $i^{th}$ party during an execution of $\Pi$ on $\bar{x}$, denoted by $V^{\Pi}_i$, is $(x_i,coins,M_i)$, where $coins$ represents the outcome of the $i^\text{th}$ party’s internal coin toss, and $M_i$ represents the set of messages it has received. For coalition $I = \{i_1,\dots,i_t \} \subseteq \{1,\dots,o \}$ of parties, the view  $V^{\Pi}_I(\bar{x})$ of the coalition during an execution of $\Pi$ is defined as
\begin{equation}
    V^{\Pi}_I(\bar{x}) = \big(I,V^{\Pi}_{i_1}(\bar{x}),\dots,V^{\Pi}_{i_t}(\bar{x})\big).
\end{equation}
\end{definition}

This means that the view of the party is all its accessible information and the view $V^{\Pi}_I(\bar{x})$ of the coalition $I$ is the union of all the views of coalition parties.
%

\begin{definition}
\label{def:semihonest}
\textbf{(Multi-party Privacy w.r.t. Semi-honest Behavior)}
Let $f:\mathbb{R}^o\mapsto \mathbb{R}^o$ be a deterministic polynomial-time function and $\Pi$ a multi-party protocol computing $f(\bar{x})$, where $\bar{x} \in \mathbb{R}^o$. For a coalition $I = \{i_1,\dots,i_t \} \subseteq \{1,\dots,o \}$ of parties, we have $\bar{x}_I = (x_{i_1},\dots,x_{i_t})$ and $f_I(\bar{x}) = \big(f_{i_1}(\bar{x}),\dots,f_{i_t}(\bar{x})\big)$. We say that $\Pi$ computes $f(\bar{x})$ privately if 
\begin{itemize}
    \item there exists a probabilistic polynomial time algorithm, denoted by simulator $S$, such that for every $I \subseteq \{1,\dots,o \}$ \citep[p.696]{conf:cryptoref}
\begin{equation}
S\big(\bar{x}_I,f_I(\bar{x}) \big)  \stackrel{c}{\equiv} V^{\Pi}_I(\bar{x}),
\end{equation}
    \item the input and output of the coalition cannot be used to infer extra private information. 
\end{itemize}
\end{definition}
Put differently, a protocol privately computes $f(\bar{x})$ if whatever can be obtained from a party’s view of a (semi-honest) execution could be essentially obtained from the input and output available to that party \citep[p. 620]{conf:cryptoref}. Also, the inputs and outputs of the coalition cannot be used to infer extra private information. Thus, our privacy proofs will always consist of the two parts of Definition \ref{def:semihonest}. 

\section{Problem Setup} \label{sec:define}





Next, let us introduce some entities for our problem setups visualized in Figures \ref{fig:setup} and \ref{fig:setupGrp}.

\begin{itemize}
\item \textbf{Plant $\mathbb{T}$:} A passive entity whose set of possible states needs to be estimated. 
We consider discrete-time linear systems with bounded noise, specifically
%
\begin{align}
\begin{split}
 x_{k+1} &= F x_k + n_k\\
y_{i,k} &= H_{i,k} x_k + v_{i,k},
\end{split}
\label{eq:sysmodel}
\end{align}
where $x_k \in \mathbb{R}^{n}$ is the state at time $k \in \mathbb{N}$, $y_{i,k} \in \mathbb{R}^{p}$ denotes the measurement observed at sensor $i$, $F$ is the process matrix, $H_{i,k}$ stands for the measurement matrix, $n_k \in \mathcal{Q}_{k}$ is the process noise bounded by process noise zonotope $\mathcal{Q}_{k}=\zono{0,Q_k }$, and $v_{i,k} \in \mathcal{R}_{k}$ is the measurement noise bounded by measurement noise zonotope $\mathcal{R}_{k}= \zono{0,\text{diag}([r_{1,k}, \dots,r_{m,k}])}$. All vectors and matrices are real-valued and have proper dimensions.



\item \textbf{Sensor $\mathbb{S}_i$:} Entity with index $i$ that provides private measurements. Its owner does not trust other active entities. 

\item \textbf{Aggregator $\mathbb{A}$ (or Cloud):} An untrusted party which has reasonable computational power. 
It executes the proposed private set-based estimation protocols over encrypted sensor information. 
\item \textbf{Query Node $\mathbb{Q}$:} An untrusted party that has a known public key \textit{pk} and a hidden private key \textit{sk}. The query node is the only node that is entitled to know the set of states of the plant $\mathbb{T}$. It might be the plant $\mathbb{T}$, but can also be any other entity other than the aggregator $\mathbb{A}$ (in order to preserve privacy).

\item \textbf{Manager $\mathbb{M}^{(j)}$:} Entity with index $j$ which estimates the state for a group of sensors and handles communication with other entities.

\item \textbf{Sensor Group $\mathbb{G}^{(j)}$:} Entity with index $j$ which consists of $m_j$ sensors  $\mathbb{S}_i^{(j)}$, $i\in\{1,...,m_j\}$, and one manager $\mathbb{M}^{(j)}$ owned by one organization. All sensors within a group trust each other and do not trust other entities. Each sensor group aims to keep its measurements and estimates private from other groups and parties. 
\end{itemize}

We provide the following definitions which are essential for set-based estimation:
\begin{definition}  \textbf{(Set-based Estimator)} Given system \eqref{eq:sysmodel} with initial state $x_0 \in \zono{c_0,G_0}$, the set-based estimator aims to find the corrected state set $\bar{\mathcal{S}}_{i,k}$ with state containment guarantees at each time step $k$, i.e., $\forall k: x_k \in \bar{\mathcal{S}}_{i,k}$.
\end{definition}
With $x_0 \in \zono{c_0,G_0}$, the predicted state set $\hat{\mathcal{S}}_{i,k}$, i.e., the set of all possible state values, is, according to \eqref{eq:sysmodel}, given by
\begin{equation}
\hat{\mathcal{S}}_{i,k}= F \hat{\mathcal{S}}_{i,k-1} \boxplus \mathcal{Q}_{i}. 
\end{equation}
For a given measurement $y_{i,k}$, the measurement state set $\mathcal{P}_{i,k}$ is the set of all possible state values satisfying the strip equation, i.e.,
\begin{equation}
    \mathcal{P}_{i,k} = \setdef[x]{\abs{ H_{i,k} x - y_{i,k}} \leq r_{i,k} }. \label{eq:strip}
\end{equation}
Where convenient, we will use the shorthand ${\mathcal{P}_{i,k}=\zono{y_{i,k},H_{i,k},r_{i,k}}}$ for a strip.
The corrected state set $\bar{\mathcal{S}}_{i,k}$ is then the over-approxima-tion of the intersection between $\hat{\mathcal{S}}_{i,k}$ and $\mathcal{P}_{i,k}$, specifically
\begin{equation}
    \bar{\mathcal{S}}_{i,k} \supseteq  \big( \hat{\mathcal{S}}_{i,k} \cap \mathcal{P}_{i,k} \big). 
\end{equation}

\begin{figure*}
    \centering
    \includegraphics[width=0.8\textwidth]{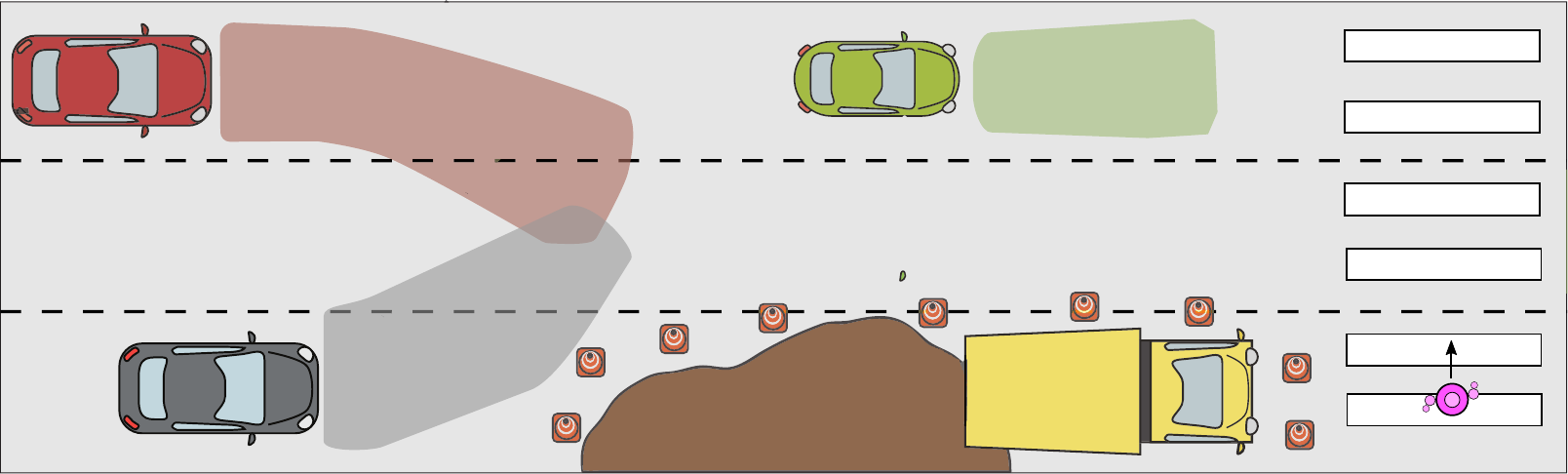}
    \caption{Highway scenario with construction work as possible application for Problems \ref{prob1} and \ref{prob2}. Reachable sets for each car are shown in their respective, transparent color \citep[adapted]{KochdumperAROC}.}
    \label{fig:highway_construction}
\end{figure*}
We aim to find solutions for the following two problems:  

\begin{problem}\label{prob1}
We want to estimate the set of possible state values of plant $\mathbb{T}$ while ensuring that measurements are private to the sensor nodes $\mathbb{S}_1, \dots ,\mathbb{S}_m$, $m\in\mathbb{N}^+$, and the estimated set is private to the query node $\mathbb{Q}$.  
\label{pb:sen}
\end{problem}


\begin{problem}\label{prob2}
We want to estimate a set of all possible state values of the plant $\mathbb{T}$ while ensuring that measurements and internally estimated sets are private to the sensor groups $\mathbb{G}_1, \dots, \mathbb{G}_d$, $d\in\mathbb{N}^+$, and the estimated set is private to the query node $\mathbb{Q}$.
\label{pb:sengrp}
\end{problem}

To illustrate the practical relevance of Problems \ref{pb:sen} and \ref{pb:sengrp}, consider the following scenario:

\begin{example}
To avoid collisions between traffic participants in a typical highway scenario (see Figure~\ref{fig:highway_construction}), each participant aims to perceive
and comprehend a traffic situation by predicting the intent of
vehicles and road users. This can be done by
computing and sharing the reachable sets of all other traffic participants, known as shared situation awareness \citep{conf:vandana}. However, computing these sets is not always possible due to computational constraints or having a participant in an occluded area from the perspective of others (see the pedestrian in Figure~\ref{fig:highway_construction}). The different entities are the following: The plant is the combination of different car dynamics communicated in an initial phase, the sensors measure the distance between the traffic participants, the cloud is the aggregator, and the street management unit that guarantees participants' safety is the query node, which aims to compute the estimated set of the position of each participant.
A possible solution hereby is to let the cloud compute reachable sets (and possible intersections thereof) while preserving the privacy of each participant (Problem~\ref{pb:sen}). Specific future scenarios may contain a car platoon trusting its participants but not other platoons (Problem~\ref{pb:sengrp}).
\end{example}

\begin{figure*}
\begin{subfigure}{.5\textwidth}
  \centering
 \includegraphics{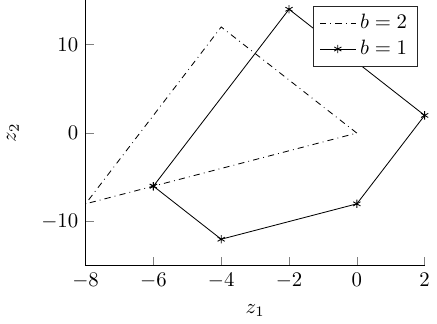}
  \caption{Two constrained zonotopes with different $b$ values.}
  \label{fig:constzono}
\end{subfigure}%
\begin{subfigure}{.5\textwidth}
  \centering
  \includegraphics{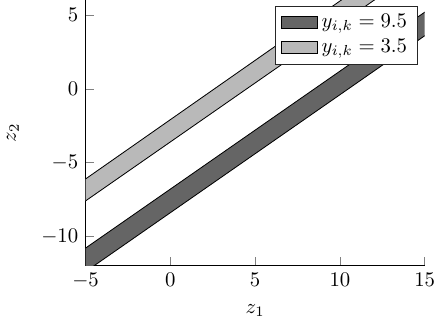}
  \caption{Two strips with different $y_{i,k}$ values. }
  \label{fig:strips-diff}
\end{subfigure}
\caption{Changing the effective parameter of constrained zonotope and strips.}
\label{fig:stripszono}
\vspace{-4mm}
\end{figure*}


For both problems it is required to guarantee computational security during the estimation process. The query node $\mathbb{Q}$ is interested in finding the set of all possible state values of plant $\mathbb{T}$ in both problems. We should note that the group manager locally estimates over unencrypted data in Problem \ref{pb:sengrp}, which is not the case for Problem \ref{pb:sen} (no group manager). If we consider the group of one sensor, there is still a need for a group manager to perform the local estimation over unencrypted data, so that Problem \ref{pb:sen} is not a special case of Problem \ref{pb:sengrp}. 

To set our privacy goals, we must first define the following coalitions that the attacker can perform for Problem \ref{pb:sen}:

\begin{definition}\label{def:sencoal}
\textbf{(Sensor Coalition)} A sensor colludes with up to $t-1$ other sensors in Problem \ref{pb:sen} by exchanging their private measurements and cryptographic private keys, constituting a sensor coalition. The coalition aims to retrieve the private information of the non-participating sensors and the query node. 
\end{definition}
\begin{definition}\label{def:aggcoal}
\textbf{(Cloud Coalition)} The aggregator $\mathbb{A}$ colludes with up to $t$ sensors in Problem \ref{pb:sen} by exchanging their private values, cryptographic private keys, and intermediate results, constituting the aggregator coalition. The coalition aims to retrieve the private information of the non-participating sensors and query node.
\end{definition}
\begin{definition} \label{def:quecoal}
\textbf{(Query Coalition)} The query node $\mathbb{Q}$ colludes with up to $t$ sensors in Problem \ref{pb:sen} by exchanging their private values, cryptographic private keys, and the final decrypted outcome of the estimation protocol, constituting the query coalition. The coalition aims to retrieve the private information of the non-participating sensors.
\end{definition}


The same definitions hold for Problem \ref{pb:sengrp} by considering sensor groups instead of sensors. We consider semi-honest parties \citep{conf:proloc} following the protocol properly, with the exception that they keep a record of all its intermediate computations to infer extra information. This paper aims to solve Problems \ref{pb:sen} and \ref{pb:sengrp} by proposing multiple secure multi-party computation protocols. These protocols should guarantee computational privacy against the aforementioned coalitions. The privacy goals are based on the concept of computational indistinguishability, which is presented next, along with the formal definition of multi-party privacy with respect to semi-honest behavior while considering coalitions.

We state and summarize the assumptions of this work subsequently:
\begin{assumption}
We assume that both process noise $n_k$ and measurement noise $v_{i,k}$ are bounded by a zonotope, i.e., $n_k \in \mathcal{Q}_{k}=\zono{0,Q_k}$, and $v_{i,k} \in \mathcal{R}_{k}= \zono{0,\text{diag}([r_{1,k}, \dots,r_{m,k}])}$.
\end{assumption}
Furthermore, we make the following assumption for the attacker's ability:
\begin{assumption}
The attacker can form sensor, aggregator, or query coalitions (see Definitions 7, 8 and 9).  
\end{assumption}
It is worth mentioning that we exclude an aggregator-query coalition, which is a common assumption in homomorphic encryption; see \citep{conf:cloutdquadratic}.

\section{Encrypted Set of States} \label{sec:propsolu}


The aforementioned set representations are deliberately chosen such that they can be used with the Paillier cryptosystem and do not reveal critical information about the measurements and estimates. More specifically, we propose using zonotopes, constrained zonotopes, and strips as set representations in privacy-preserving set-based estimation: 

\begin{enumerate}[leftmargin=12pt,noitemsep,partopsep=0pt,topsep=0pt,parsep=0pt]
\itemsep0em
    \item Zonotopes: 
    We homomorphically encrypt the center and reveal the generator matrix, thus hiding the position of the zonotope and only revealing the estimation uncertainty (zonotope shape) described by the generator matrix.
    \item Constrained zonotopes: 
      To protect the estimation uncertainty (more privacy) while introducing extra computation overhead, we use constrained zonotopes instead of zonotopes. Changing $b$ of a constrained zonotope $\zono{c,G,A,b}$ (see Definition~\ref{df:contzono}) changes both its position and shape as shown in Figure~\ref{fig:constzono}. We propose to encrypt the vectors $c$ and $b$ and thus encrypt both position and shape of the set. 
\item Strips: For a strip given by  \eqref{eq:strip},  we encrypt $y_{i,k}$, and reveal $H_{i,k}$ and $r_{i,k}$. Encrypting $y_{i,k}$ preserves the privacy of the strip position as shown in Figure~\ref{fig:strips-diff} for two strips with $H_{i,k} = [-1.25, 1]$, $r_{i,k}=1$, and $y_{i,k} \in \setdef{3.5,9.5}$. 
\end{enumerate}


The chosen selective encryption will allow us to decouple the computation of public information from private information, as we will show later. We propose two protocols to solve Problems \ref{pb:sen} and \ref{pb:sengrp} while preserving the mentioned privacy goals. Two variants of the proposed protocols solve the problems using zonotopes while revealing the estimation uncertainty. The other two variants solve the problems using constrained zonotopes while preserving the uncertainty around the estimates. We start by discussing the protocol solving Problem \ref{pb:sen}.

\section{Private Estimation Using Distributed Sensors} \label{sec:alg1}
In this section, we introduce a protocol for estimating the set of possible states using zonotopes and constrained zonotopes while achieving our privacy goals. We first describe both protocols using a general set representation and then specify the required operations for zonotopes and constrained zonotopes. The query node $\mathbb{Q}$ generates the Paillier public key \textit{pk} and private key \textit{sk} and shares the public key with other parties. It then chooses a large enough initial set of possible states, enclosing the true state according to the public information. The initial set $\hat{S}_{q,0}$ is encrypted by the query node. We add the subscript $q$ to the set notation to indicate that the set computation is done at the query node. The initial encrypted set $\enc{\hat{S}_{q,0}}$ is sent to the aggregator.

\begin{figure}[t]
    \centering
     \includegraphics[width=0.45\textwidth]{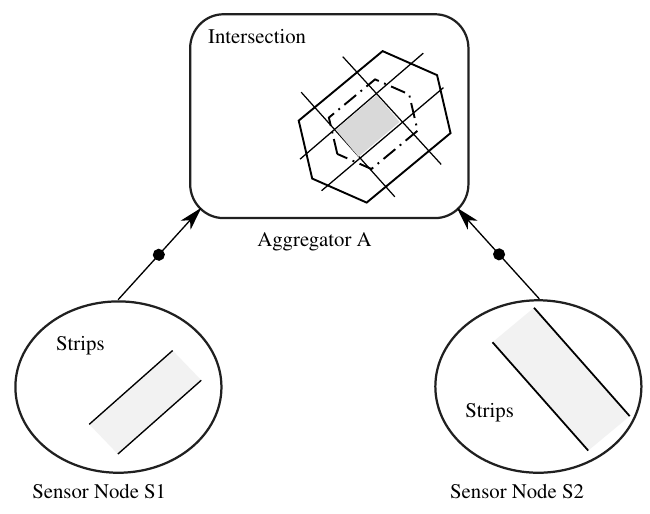}
    \caption{Overview of Protocol \eqref{alg:1} where every sensor shares an encrypted strip with the aggregator, which then privately intersects these encrypted strips with an encrypted zonotope.}
    \label{fig:prt1}
    \vspace{-5mm}
\end{figure}

Our proposed privacy-preserving approach consists of three steps: the measurement update, the time update, and sharing of the results in a continuous loop, as presented in Protocol \ref{alg:1}. More specifically, during the measurement update, the aggregator collects an encrypted strip $\enc{\mathcal{P}_{i,k}}$ from each sensor node $i$ at step $k$, as shown in Figure \ref{fig:prt1}. The family of encrypted strips (measurements) is intersected in the encrypted domain with the predicted reachable set at the aggregator (indicated by subscript $a$) -- initially, it is the initial encrypted set $\enc{\hat{S}_{q,0}}$ sent by the query node -- to obtain the encrypted corrected set $\enc{\bar{\mathcal{S}}_{a,k}}$, shown in Figure \ref{fig:prt1}. Finally, the aggregator performs the time update and sends the encrypted corrected set $\enc{\hat{\mathcal{S}}_{a,k}}$, after decreasing its order, to the query node, which decrypts the result for each time step $k$.

\begin{algorithm}[t!]
\floatname{algorithm}{Protocol}
\caption{Private Estimation using Distributed Sensors}
\label{alg:1}
\begin{algorithmic}
\State The query node $\mathbb{Q}$ encrypts the initial set $\enc{\hat{\mathcal{S}}_{q,0}}$ and sends it to the aggregator node to have $\enc{\hat{\mathcal{S}}_{a,0}}=\enc{\hat{\mathcal{S}}_{q,0}}$. At every time instant $k$, every sensor node shares an encrypted strip $\enc{\mathcal{P}_{i,k}}=\langle\enc{y_{i,k}},H_k,r_{i,k}\rangle$ with the aggregator which executes the following steps: 
\State \textbf{Step 1}: Measurement update at the aggregator:
\begin{align}
\enc{\bar{\mathcal{S}}_{a,k}} = \enc{\hat{\mathcal{S}}_{a,k-1}} \cap \enc{\mathcal{P}_{1,k}} \cap \dots \cap \enc{\mathcal{P}_{m,k}}
\end{align} 
\State \textbf{Step 2}: Time update at the aggregator:
\begin{align}
\enc{\tilde{\mathcal{S}}_{a,k}} =& F \enc{\bar{\mathcal{S}}_{a,k}} \boxplus \mathcal{Q}_{k} \label{eq:ptc1_update_set}\\
\enc{\hat{\mathcal{S}}_{a,k}}  =&   \downarrow_q \enc{\tilde{\mathcal{S}}_{a,k}}\label{eq:ptc1_reduce_set} 
\end{align}
\State  \textbf{Step 3}: The aggregator sends the encrypted set $\enc{\hat{\mathcal{S}}_{a,k}}$ to the query node which decrypts the result for each time step $k$.
\end{algorithmic}
\end{algorithm}

We start by describing the required operations for the zonotopic case. The intersection between zonotopes and a family of strips can be performed according to \citep{conf:stripzono} and is summarized in the following lemma:

\begin{lemma}(\citep[Prop.1]{conf:stripzono})
\label{lem:stripzono}
The intersection $\hat{\mathcal{Z}}_{k-1} \cap \mathcal{P}_{1,k} \cap \dots \cap \mathcal{P}_{m,k}$ of a zonotope $\hat{\mathcal{Z}}_{k-1}=  \langle \hat{c}_{k-1},\hat{G}_{k-1}  \rangle $ and the family of $m$ strips $\mathcal{P}_{j,k}=\zono{y_{i,k},H_{i,k},r_{i,k}}$ in \eqref{eq:strip}, $\forall j \in \mathcal{N}$, $\abs{\mathcal{N}} = m$, is overapproximated by a zonotope $\bar{\mathcal{Z}}_{k}=\zono{  \bar{c}_{k},\bar{G}_{k}}$, where $\lambda_{j,k} \in \R^{n \times p}$ is the design parameter, and 
\begin{align}
  \bar{c}_{k} &=  \hat{c}_{k-1} + \sum\limits_{j\in\mathcal{N}} \lambda_{j,k}(y_{j,k} - H_{j,k} \hat{c}_{k-1}), \label{eq:C_lambda}\\
  \bar{G}_{k} &= \Big[ (I - \sum\limits_{j\in\mathcal{N}} \lambda_{j,k} H_{j,k} ) \hat{G}_{k-1}, \lambda_{1,k} r_{1,k},\dots,\lambda_{m,k} r_{m,k} \Big]. 
  \label{eq:G_lambda}
 \end{align}
  \end{lemma}

The factor $\lambda_{j,k} \in \mathbb{R}^{n \times p}$ is a degree of freedom in Lemma~\ref{lem:stripzono} which we use to maximize the tightness of the intersection over-approximation. Thus, we want to find $\Lambda_{k} =\begin{bmatrix} \lambda_{1,k},  \ldots, \lambda_{m,k} \end{bmatrix}$ that decreases the uncertainty around the estimates. We achieve this by computing the $\Lambda_{k}$ that decreases the Frobenius norm of the generator matrix $\bar{G}_{k}$ in \eqref{eq:G_lambda} \citep{conf:gainOpt}. 

During the time update step, the aggregator computes the time evolution of the estimated encrypted zonotope according to \eqref{eq:ptc1_update_set} and \eqref{eq:ptc1_reduce_set}, i.e.,
\begin{align}
    \hat{c}_{a,k} =&  F \bar{c}_{a,k}, \label{eq:chatzonotimeupdate}\\
    \tilde{G}_{a,k} =& [F \bar{G}_{a,k},  Q_k], \ \ \ 
    \hat{G}_{a,k}  =   \downarrow_q \tilde{G}_{a,k}. \label{eq:Gtildehatzonotimeupdate}
\end{align}

Decreasing the order of the generator matrix, denoted by $\downarrow_q$, is done according to \citep{conf:girard}, which can be done over encrypted set as the generator matrix is revealed. 

The generators do not participate in determining the position of the zonotope. Thus, it is sufficient to process over encrypted zonotope centers, as clarified in Section \ref{sec:propsolu}. Given the nature of the intersection between the strips and a zonotope in Lemma \ref{lem:stripzono}, the operations in the encrypted domain are decoupled from the plaintext domain computations. Note that the Paillier properties in \eqref{eq:pailadd} and \eqref{eq:pailmul} allow one to process \eqref{eq:C_lambda} over the encrypted center $\enc{\bar{c}_{k}}$ and measurement $\enc{y_{j,k}}$ in Protocol~\ref{alg:1}. This protocol computes \eqref{eq:C_lambda} in the encrypted domain and \eqref{eq:G_lambda} in the unencrypted domain, where we also compute $\Lambda_{k}$. More specifically, we operate over encrypted centers and measurements as follows:
\begin{align}
      \enc{\bar{c}_{k}} =&  \enc{\hat{c}_{k-1}} \oplus \sum\limits_{j\in\mathcal{N}} \lambda_{j,k}(\enc{y_{j,k}} \ominus H_{j,k} \enc{\hat{c}_{k-1}}), \\
      \enc{\hat{c}_{a,k}} =&  F \enc{\bar{c}_{a,k}}.
\end{align}
The generators are in the unencrypted domain in all our algorithms. The Minkowski sum in \eqref{eq:ptc1_update_set} is computed over encrypted centers and unencrypted generators.

The next theorem summarizes the privacy of Protocol \ref{alg:1} against different coalitions in Definitions \ref{def:sencoal}, \ref{def:quecoal}, and \ref{def:aggcoal} when we use zonotopes and strips as sets.
 
 \begin{theorem}
 \label{th:alg1pricol}
 Protocol \ref{alg:1} solves Problem \ref{pb:sen} using encrypted zonotopes while revealing the shape of the estimated zonotope and achieving privacy against 
 \begin{itemize}
     \item sensor coalitions,
     \item cloud coalitions,
     \item query coalitions if $m_r p > n$, where $m_r$ is the number of non-colluding sensors, $p$ the measurement size, $n$ the size of the state.
 \end{itemize}

 \end{theorem}
 
 The proof is detailed in the Appendix. To overcome the information leakage in case of query coalitions when $m_r p \leq n$, we propose a slight modification by keeping the strip parameter $r_{i,k}$ private to the sensor and aggregator. Then, the aggregator swaps the columns of the generator matrix $\hat{G}_{a,k}$ before sending it to the query node. Swapping the generator columns produces the same estimated zonotope, but preserves privacy by preventing the coalition from computing $\Lambda_{k}$ and thus also prevents the extraction of the center $\enc{\hat{c}_{a,k}}$.

Next, we present the required operations using constrained zonotopes. The following theorem shows the intersection in the unencrypted domain.

 \begin{lemma}
 \label{lm:conststrips}
 
 The intersection $\hat{\mathcal{C}}_k \cap \mathcal{P}_{1,k} \cap \dots \cap \mathcal{P}_{m,k}$ of a constrained zonotope $\hat{\mathcal{C}}_k=\langle \hat{c}_k,\hat{G}_k,\hat{A}_k,\hat{b}_k \rangle$ and the family of $m$ strips $\mathcal{P}_{j,k}=\zono{y_{i,k},H_{i,k},r_{i,k}}$ in \eqref{eq:strip}, $\forall j \in \mathcal{N}$, $\abs{\mathcal{N}}=m$, is a constrained zonotope $\bar{\mathcal{C}}_k=\zono{\bar{c}_k,\bar{G}_k,\bar{A}_k,\bar{b}_k}$ where $\lambda_{j,k} \in \R^{n \times p}$ is a degree of freedom and
%
 \begin{align}
  \bar{c}_k &=  \hat{c}_k + \sum\limits_{j\in\mathcal{N}} \lambda_{j,k}(y_{j,k} - H_{j,k} \hat{c}_k ),\\
  \bar{G}_k &= \Big[ (I - \sum\limits_{j\in\mathcal{N}} \lambda_{j,k} H_{j,k} ) \hat{G}_k, \lambda_{1,k} r_{1,k},\dots,\lambda_{m,k} r_{m,k} \Big], 
  \label{eq:G_lambdapr3}\\
  \bar{A}_k &= \begin{bmatrix}\hat{A}_k &0& 0&\dots& 0 \\ H_{1,k} \hat{G}_k &-r_{1,k}& 0 &\dots& 0 \\ H_{2,k} \hat{G}_k &0&-r_{2,k} &\dots& 0 \\ 
\vdots &\vdots  & &\ddots &\label{eq:Abarconstzono}\\
H_{m,k} \hat{G}_k & 0 & 0 & \dots & -r_{m,k} \end{bmatrix},\\ 
\bar{b}_k &= \begin{bmatrix}\hat{b}_k \\ y_{1,k} - H_{1,k} \hat{c}_k\\ y_{2,k} - H_{2} \hat{c}_k \\ \vdots \\ y_{m,k} - H_{m,k} \hat{c}_k \end{bmatrix}. \label{eq:bbarconstzono}
 \end{align}
 \end{lemma}
 \begin{proof}
 The results can be obtained by applying an intersection from \citep[Prop. 1]{conf:const_zono} and then adding a degree of freedom from \citep[Prop. 5]{conf:const_zono}. 
 \end{proof}

During the measurement update in Protocol \ref{alg:1} when using constrained zonotopes, the aggregator performs the proposed intersection over the encrypted center $\enc{\hat{c}_{a,k-1}}$ and the encrypted constraint shift $\enc{\hat{b}_{a,k-1}}$. Note that various combinations of zonotopes and constraints can represent the same constrained zonotope. Here, we exploit the additional degree of freedom in $\Lambda_{k}$ and choose it at random in our protocol, which improves privacy, as discussed in the Appendix. Next, the aggregator propagates the sets forward in time according to \eqref{eq:sysmodel}, and then reduces the order of the given set \citep{conf:const_zono}, i.e.,
\begin{align}
\hat{c}_{a,k} &= F \bar{c}_{a,k}, \ \ 
\tilde{G}_{a,k} = [F \bar{G}_{a,k},  Q_k], \ \ 
\hat{b}_{a,k} = \bar{b}_{a,k}, \label{eq:chatGtildbhatpr3}\\
 \{\hat{G}_{a,k}&, \hat{A}_{a,k}\}  =   \downarrow_q \{\tilde{G}_{a,k}, \bar{A}_{a,k}\} \label{eq:GAhatpr3}.
\end{align}

The privacy of Protocol \ref{alg:1} against different coalitions in Definitions \ref{def:sencoal}, \ref{def:aggcoal}, and  \ref{def:quecoal} is summarized in the following theorem. 

 \begin{theorem}
 \label{th:alg3pricol}
 Protocol \ref{alg:1} solves Problem \ref{pb:sen} using encrypted constrained zonotopes while protecting the shape of the estimated set and achieves privacy against 
 \begin{itemize}
    \item sensor coalitions,
    \item cloud coalitions,
    \item query coalitions.
 \end{itemize}

 \end{theorem}
 
 

The proof is detailed in the Appendix. After presenting our constrained zonotopic privacy-preserving protocol for Problem~\ref{pb:sen}, we move on to the privacy-preserving protocol for Problem~\ref{pb:sengrp}.

\section{Private Estimation Using Sensor Groups} \label{sec:alg2}

\begin{figure}[t]
  \centering
 \includegraphics[scale=0.59]{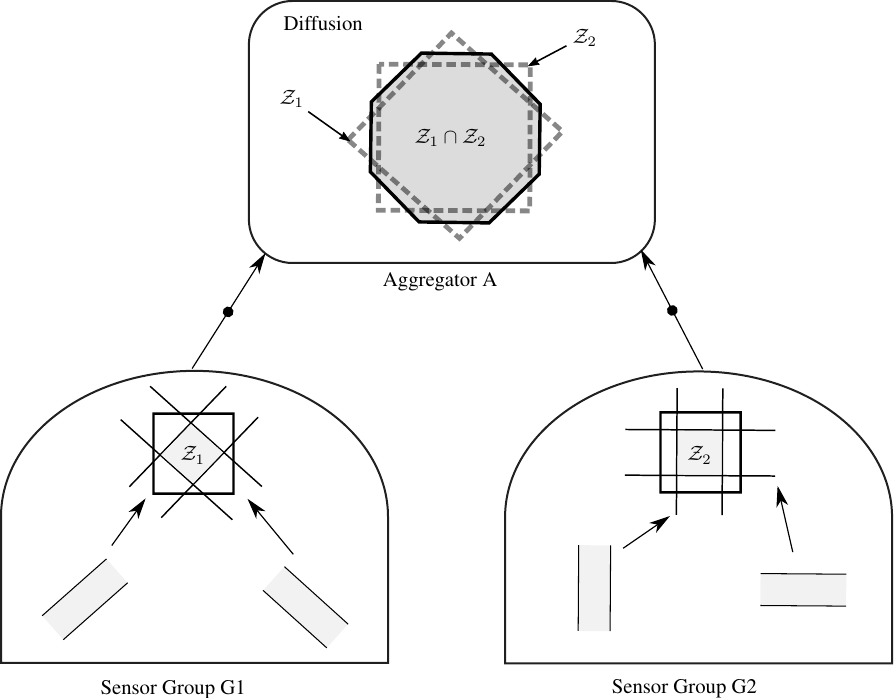}
 \vspace{-3mm}
  \caption{Overview of Protocol \ref{alg:2} where every sensor group computes the intersection between a zonotope and its strips. The aggregator then computes the intersection over the encrypted zonotopes.}
  \label{fig:prt2}
\vspace{-4mm}
\end{figure}
We provide a privacy-preserving protocol for Problem \ref{pb:sengrp} in Protocol \ref{alg:2}, which is represented graphically in Figure \ref{fig:prt2} for the zonotopic case. Each sensor $j$ within group $i$ is participating with a strip (measurement) set $\mathcal{P}_{j,k}^{(i)}$ at each time step $k$. All strips are collected within the group $i$ and are then intersected with the previously estimated set $\hat{\mathcal{S}}_{a,k-1}$ in \eqref{eq:setstripintersengroup} in the unencrypted domain, as the group participants trust each other and the plaintext execution is faster than the encrypted domain execution. The owner of each sensor group encrypts the resulting set $\bar{\mathcal{S}}^{(i)}_k$ and sends it to the aggregator, which in turn computes the intersection of all received encrypted sets in the encrypted domain in \eqref{eq:setdiff}. Next, the aggregator performs the time update. Finally, the aggregator submits $\enc{\hat{\mathcal{S}}_{a,k}}$ to the query node, which decrypts the result and sends it to each sensor group. 


We first describe the required operations for zonotopes and then proceed with constrained zonotopes. In the measurement update step, the intersection between a zonotope and a family of strips is performed as described in Lemma \ref{lem:stripzono}. Unlike Protocol \ref{alg:1}, where we perform the aforementioned intersection at the aggregator in the encrypted domain, we now intersect at the sensor group level in the unencrypted domain since each sensor trusts all other sensors from the same group. Different methods exist in the literature for the zonotope intersection required during the diffusion update. Here, we picked our previously proposed intersection method described in \citep{conf:dis-diff}, which fits the homomorphic computations, as summarized in the following lemma. 

\begin{lemma}(\citep[Th.2]{conf:dis-diff})
 \label{lem:diff}
The intersection $\bar{\mathcal{Z}}_{1,k} \cap \dots \cap \bar{\mathcal{Z}}_{d,k}$ between $d$ zonotopes $\bar{\mathcal{Z}}_{j,k}=\zono{\bar{c}_{j,k}, \bar{G}_{j,k} }$, $j\in \{1,\dots,d\}$, can be over-approximated using the zonotope $\grave{\mathcal{Z}}_{k}=\langle \grave{c}_{k},\grave{G}_{k} \rangle $ given by
 \begin{eqnarray}
\grave{c}_{k}&=&\frac{1}{\sum\limits_{j \in \mathcal{N}} w_{j,k}}\sum\limits_{j \in \mathcal{N}} w_{j,k} \bar{c}_{j,k},\label{eq:cdiff}\\
\grave{G}_{k} &=& \frac{1}{\sum\limits_{j \in \mathcal{N}}w_{j,k}}[w_{1,k}\bar{G}_{1,k} ,...,w_{d,k}\bar{G}_{d,k}], \label{eq:gdiff}
 \end{eqnarray}
 where the weights $w_{j,k}$ are chosen such that $\sum \limits_{j \in \mathcal{N}} w_{j,k} \neq 0$.
 \end{lemma}
Let $\mathfrak{w}_{k}=[w_{1,k},\dots,w_{d,k}]$, where $d$ is the number of sensor groups. Ideally, $\mathfrak{w}_{k}$ is chosen such that the size of the zonotope $\grave{\mathcal{Z}}_{k}=\langle \grave{c}_{k},\grave{G}_{k} \rangle$ is minimized. The size of the zonotope appears in the unencrypted generator matrix due to the selective encryption, and can be replaced by the Frobenius norm of the generator matrix. The next theorem summarizes the privacy features of the protocol against different coalitions in Definitions \ref{def:sencoal}, \ref{def:quecoal}, and \ref{def:aggcoal}.

\begin{algorithm}[t]
\floatname{algorithm}{Protocol}
\caption{Private Estimation using Sensor Groups}
\label{alg:2}
\begin{algorithmic}
\State The query node $\mathbb{Q}$ sends the initial set to each sensor group $i\in\setdef{1,...,m}$, and the aggregator node $\mathbb{A}$. For each sensor group $i$, $m_i$ strips $\enc{\mathcal{P}^{(i)}_{j,k}}=\zono{\enc{y^{(i)}_{j,k}},H_{k},r^{(i)}_{j,k}}$, $j\in\setdef{1,...,m_i}$, are available. At every time instant $k$, the following steps are executed:
\State \textbf{Step 1}: Measurement update at each sensor group $i$:
\begin{align}
\bar{\mathcal{S}}^{(i)}_k &= \hat{\mathcal{S}}_{a,k-1} \cap \mathcal{P}^{(i)}_{1,k} \cap \dots \cap \mathcal{P}^{(i)}_{m_i,k} 
\label{eq:setstripintersengroup}
\end{align}
\State \textbf{Step 2}: Diffusion update at the aggregator:
\begin{align}
\enc{\grave{\mathcal{S}}_{a,k}} &=  \enc{\bar{\mathcal{S}}^{(1)}_k} \cap \dots \cap \enc{\bar{\mathcal{S}}^{(d)}_k}
\label{eq:setdiff}
\end{align}
\State \textbf{Step 3}: Time update at the aggregator:
\begin{align}
\enc{\tilde{\mathcal{S}}_{a,k}} =& F \enc{\grave{\mathcal{S}}_{a,k}} \boxplus \mathcal{Q}_{k} \\
\enc{\hat{\mathcal{S}}_{a,k}}  =&   \downarrow_q \enc{\tilde{\mathcal{S}}_{a,k}} 
\end{align}
\State \textbf{Step 4}: The aggregator sends the encrypted set $\enc{\hat{\mathcal{S}}_{a,k}}$ to the query node which decrypts and sends the results to the sensor groups.
\end{algorithmic}
\end{algorithm}

\begin{theorem}
 \label{th:alg2pricol}
 Protocol \ref{alg:2} solves Problem \ref{pb:sengrp} using encrypted zonotopes while revealing the shape of the estimated set and achieving privacy against 
  \begin{itemize}
     \item sensor coalitions,
     \item cloud coalitions,
     \item query coalitions if $(d_r > 1)$, where $d_r$ is the number of non-colluding sensor groups. 
 \end{itemize}
\end{theorem}

 The proof is detailed in the Appendix. In order to solve Problem \ref{pb:sengrp} without revealing the shape of the estimated set as in Theorem \ref{th:alg2pricol}, we again use constrained zonotopes. The intersection between the constrained zonotopes and strips during the measurement update of Protocol \ref{alg:1} is done according to Lemma \ref{lm:conststrips} in the unencrypted domain. Then, the intersection between the constrained zonotopes is performed, which we describe next.

\begin{lemma}
 \label{lem:diffconst}
The intersection $\bar{\mathcal{C}}_{1,k} \cap \dots \cap \bar{\mathcal{C}}_{d,k}$ between $d$ constrained zonotopes $\bar{\mathcal{C}}_{j,k}=\left\langle\bar{c}_{j,k}, \bar{G}_{j,k}, \bar{A}_{j,k}, \bar{b}_{j,k} \right\rangle$ is a constrained zonotope $\grave{\mathcal{Z}}_{k}=\zono{\grave{c}_{k},\grave{G}_{k},\grave{A}_k, \grave{b}_k }$, where
 \begin{align}
 \grave{c}_{k}  &=  \bar{c}_{1,k},\ \ \ \grave{G}_{k} = \Big[  \bar{G}_{1,k}, 0, \dots ,0 \Big],  \label{eq:cGgravpr4}\\
  \grave{A}_{k}  &= \begin{bmatrix}\bar{A}_{1,k} & 0&\dots& 0 \\ 
  0  &\bar{A}_{2,k} &\dots& 0 \\ 
\vdots  &\vdots  &\ddots &\vdots\\
0  & 0 & \dots & \bar{A}_{{d},k}  \\
\bar{G}_{1,k}  & -\bar{G}_{2,k} & \dots & 0 \\
\vdots  &\vdots  &\ddots &\vdots\\
\bar{G}_{1,k}  & 0 & \dots & -\bar{G}_{d,k} \end{bmatrix},\ \ \ 
  \grave{b}_{k}  =  \begin{bmatrix}\bar{b}_{1,k} \\ \bar{b}_{2,k} \\ \vdots \\\bar{b}_{d,k} \\  \bar{c}_{2,k}  - \bar{c}_{1,k}  \\ \vdots \\ \bar{c}_{d,k} - \bar{c}_{1,k}  \end{bmatrix}. \label{eq:Abgravpr4}
 \end{align}
 \end{lemma}
 \begin{proof}
 The lemma is the multi-strip intersection of \citep[Prop. 1]{conf:const_zono}.
 \end{proof}
 
 Unlike for zonotopes, Lemma \ref{lem:diffconst} computes the intersection exactly. Then, the time update is done according to \eqref{eq:chatGtildbhatpr3} and \eqref{eq:GAhatpr3}. The privacy of the protocol against different coalitions in Definitions \ref{def:sencoal}, \ref{def:quecoal}, and \ref{def:aggcoal} is summarized in the following theorem.
 
  \begin{theorem}
 \label{th:alg4pricol}
 Protocol \ref{alg:2} solves Problem \ref{pb:sengrp} using encrypted constrained zonotopes while protecting the shape of the estimated set and achieves privacy against 
 \begin{itemize}
    \item sensor coalitions,
    \item cloud coalitions,
    \item query coalitions.
 \end{itemize}

 \end{theorem}
 

The proof is detailed in the Appendix. In the next section, we will evaluate the presented protocols.

\section{Evaluation} \label{sec:eval}

\begin{figure*}[t!]
\centering
\begin{minipage}[t]{0.28\textwidth}%
\includegraphics[height=1.2in]{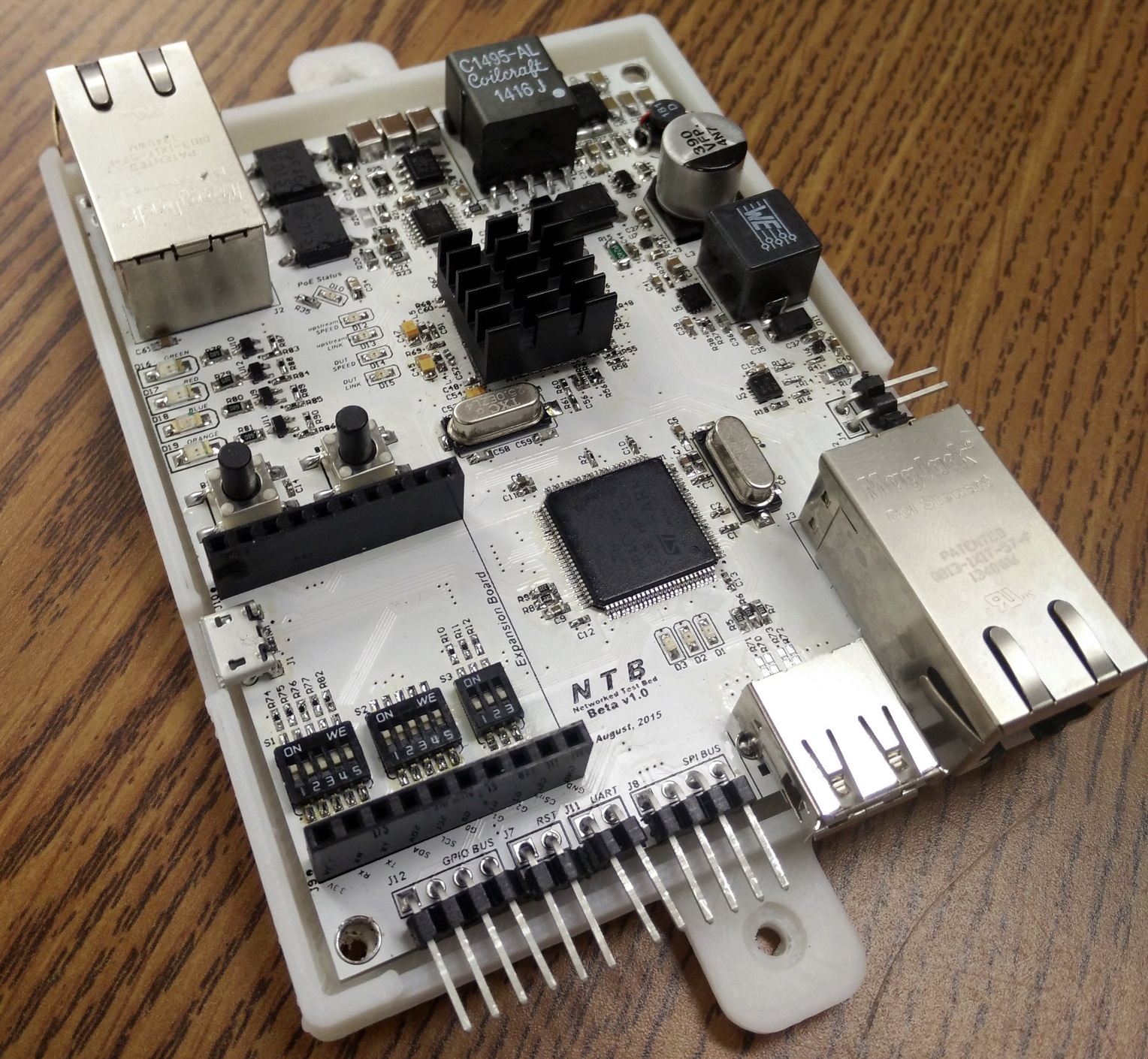}
\caption{Custom anchor with ARM Cortex M4 processor and UWB slot.}
\label{fig:ntbanon}
\end{minipage}\hspace{4mm}
\begin{minipage}[t]{0.28\textwidth}%
\centering
\includegraphics[height=1.2in]{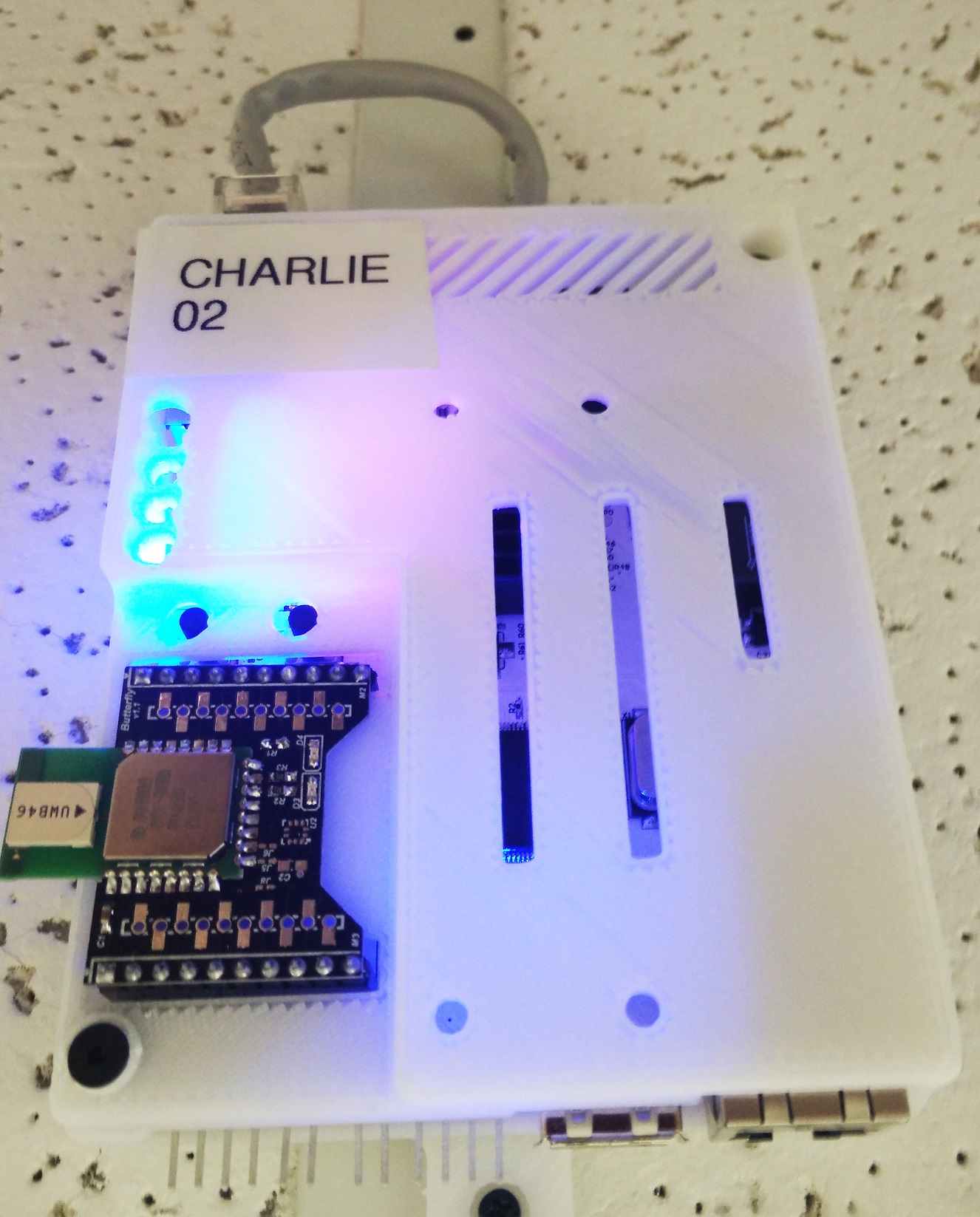}
\caption{Ceiling-mounted anchor with UWB radio in 3D-printed enclosure.}
\label{fig:ntbceil}
\end{minipage}\hspace{4mm}
\begin{minipage}[t]{0.28\textwidth}%
\centering
\includegraphics[height=1.2in]{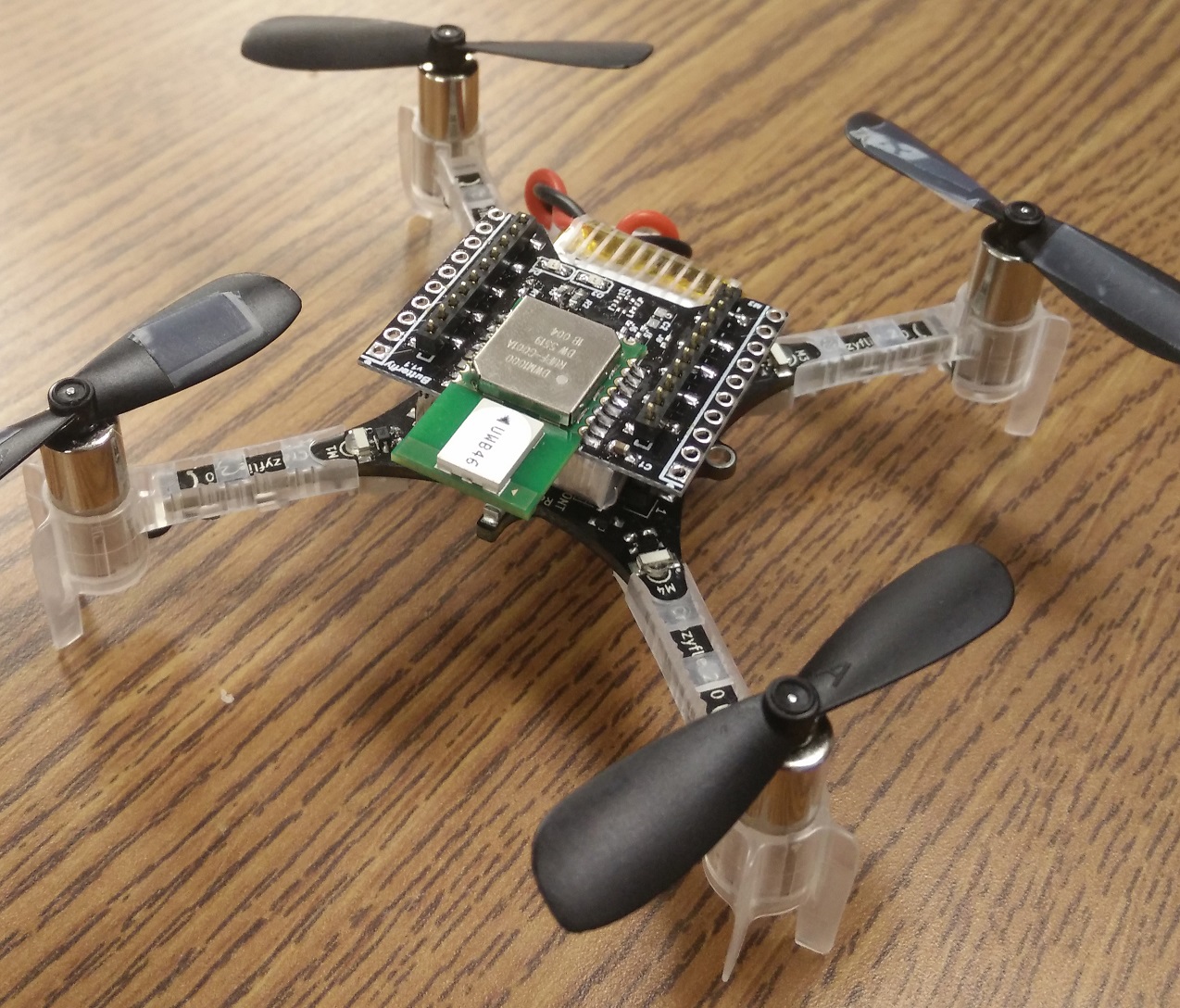}
\caption{CrazyFlie 2.0 quadrotor helicopter with UWB expansion.}
\label{fig:quad}
\end{minipage}
\vspace{-3mm}
\end{figure*}

In this section, we evaluate the proposed protocols using data from a real-world testbed. We first describe our testbed in detail and then evaluate the proposed protocols. The protocols are evaluated on a custom ultra-wideband (UWB) RF testbed based on the DecaWave DW1000 IR-UWB radio\footnote{Decawave DW1000:  \url{http://www.decawave.com/products/dw1000}}. The overall setup is the same as in \citep{conf:tersekf}. The main components of the considered testbed can be summarized as follows: 
\begin{enumerate}
    \item The motion capture system consists of eight cameras capable of performing 3D rigid body position measurements with an accuracy of less than \SI{0.5}{\milli\meter}. \item The fixed nodes each consist of a custom-built circuit board equipped with a ARM Cortex M4 processor with \SI{196}{\mega\hertz} (Figure \ref{fig:ntbanon}), powered over Ethernet and communicating via a Decawave DW1000 ultra-wideband radio (Figure \ref{fig:ntbceil}). 
    \item The battery-powered mobile node is a modified CrazyFlie 2.0 helicopter\footnote{Bitcraze CrazyFlie 2.0: \url{https://www.bitcraze.io/}} (Figure \ref{fig:quad}) and is equipped with the same DW1000 radio and ARM Cortex M4 processor.
\end{enumerate}

\begin{figure*}
    \centering
 \includegraphics[scale=1.2]{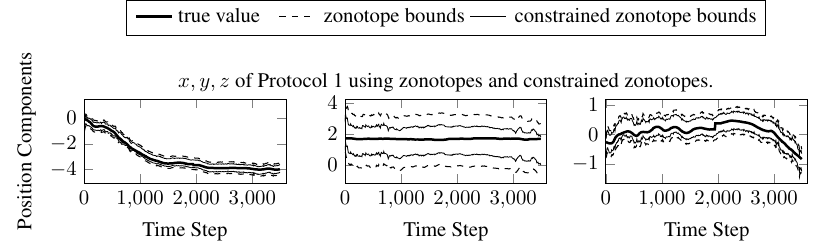}
    \vspace{-1mm}
    \caption{Ground truth, upper and lower bounds on the position components of the three-dimensional estimated states in meters of Protocol \ref{alg:1}.}
\label{fig:results1to4_protocols}
\end{figure*}

For the sake of a fair evaluation between the four variants of our two protocols, we used a collected data from the testbed and ran the four variants on the same set of measurements. We aim to estimate the set that encloses the location of the quadrotor while preserving our aforementioned privacy goals. We start with an initial set of size ($8 \times 8$ $m^2$) covering the whole localization area at the initial point (time step $k=0$). This set is then iteratively shrunk by using the received measurements and performing geometric intersections to correct the estimated set. Figure \ref{fig:results1to4_protocols} shows the true values, upper bounds, and lower bounds of the three-dimensional estimated location of the four variants of Protocol~\ref{alg:1}. We omit the results of Protocol~\ref{alg:2} as they are close to the results of Protocol~\ref{alg:1}. The upper bounds and lower bounds are obtained by converting the zonotopes and constrained zonotopes to intervals. It is worth mentioning that the result using the zonotopic case of Protocol~\ref{alg:1} is tighter than the result using the zonotopic case of Protocol~\ref{alg:2}. This is because Protocol \ref{alg:2} requires two over-approximations, namely, the intersection between every zonotope and the family of strips and the intersection of the family of zonotopes. 

\begin{figure*}[ht]
\centering
\def\scale{0.25}
\includegraphics[scale=1.2]{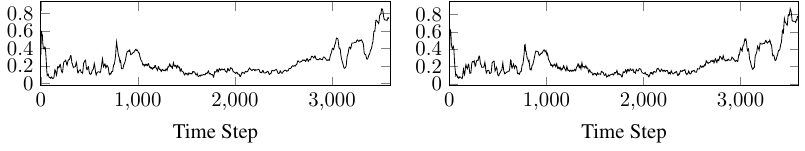}
\caption{Estimation error for Protocol \ref{alg:1} using zonotopes and constrained zonotopes.}
\label{fig:esterr}
\vspace{-3mm}
\end{figure*}

We consider the center of the estimated set to be the single-point estimate in the zonotopic case. Thus, we report the localization error with respect to the center of the zonotope. For constrained zonotopes, the reported center in the representation is the center of the original zonotope without constraints and hence can be outside of the constrained zonotope. Therefore, we compute the Chebychev center of the polytope in the constrained zonotope factor space \citep{conf:cora3}. The estimation error of the four variants is presented in Figure \ref{fig:esterr}. 

There is a trade-off between the provided privacy, the computation overhead, and the exactness of set operations. Constrained zonotopes provide more privacy, more computation overhead, and less conservatism due to the exact set operations. On the other hand, zonotopes provide less privacy due to revealing the shape of the sets, less computation overhead, and more conservative sets. The trade-off between the provided privacy and the execution time is presented in Table \ref{tab:exectime}. Keeping the shape of the estimated set private by using constrained zonotopes instead of zonotopes increases the required execution time. All computations were run on a single thread of an Intel(R) Core(TM) i7-8750 with 16 GB RAM with 1024 key size. The comparison between the size of the reachable sets appears in Figure \ref{fig:results1to4_protocols}.

\begin{table*}[tbp]
\caption{Execution Time in \SI{}{\milli\second}.}
\label{tab:exectime}
\centering
\normalsize
\begin{tabular}{c  c c c}
\toprule
   & \multicolumn{3}{c}{Entities}\\
  \cmidrule(lr){2-4} 
  & Sensor/Sensor group & Aggregator & Query\\
\midrule
Protocol \ref{alg:1} using zonotopes & 2.371   & 3.195  & 0.550 \\
Protocol \ref{alg:1} using constrained zonotopes & 2.371  & 6.632 &  14.529\\
Protocol \ref{alg:2} using zonotopes & 8.389  & 5.968 & 0.550\\
Protocol \ref{alg:2} using constrained zonotopes & 81.426 & 9.787 &  14.529\\
\bottomrule
\end{tabular}
\vspace{-4mm}
\end{table*}

%
%
%
%

\section{Conclusions} \label{sec:con}

We proposed the first privacy-preserving, set-based observers using homomorphic encryption. We considered both a traditional sensor setup as well as a scenario where trusting sensors are grouped into sensor groups. We showed that by choosing zonotopes and constrained zonotopes to represent our sets, it is possible to selectively encrypt only the critical set parameters while achieving the desired level of privacy. To prove that privacy for each protocol, the concept of computational indistinguishability was used. Finally, we evaluated our algorithms on real data from a physical test bed, which showed that the proposed protocols achieve satisfactory results while guaranteeing privacy.

One main drawback of guaranteeing privacy using homomorphic encryption is the overflow problem after a sequence of operations in the encrypted domain. To overcome the overflow limitation, we send the encrypted set to the query node each time, which decrypts the estimated set and sends the re-encrypted set back to the aggregator. This solves the overflow problem at the cost of computation and communication overhead. However, after the encrypted estimated set is sent from the aggregator to the query node, decrypting said set is not regarded as overhead since the query node is interested in the estimated set after each time step by assumption, and thus decryption is required anyway. That said, solving the overflow problem in a more efficient way is an open research problem that we leave for future work.


\section*{Acknowledgements}
We gratefully acknowledge partial financial support by the project justITSELF funded by the European Research Council (ERC) under grant agreement No 817629, the project interACT under grant agreement No 723395, and the CONCORDIA cyber security project No. 830927; these projects are funded within the EU Horizon 2020 program. This work is also supported by the Knut and Alice Wallenberg Foundation, the Swedish Foundation for Strategic Research and the Swedish Research Council.
\appendix



\section{Theorems' Proofs}
 We need to show in the following proofs that the views and simulators of the coalitions are computationally indistinguishable and that the input and output of the coalition do not leak extra private information according to Definition \ref{def:semihonest}. This will be done for each type of coalition. Showing the computational indistinguishability is done by building the views of each coalition then proving that there is an equivalent simulator that could be obtained from the input and output available to the coalition. 
  \begin{remark} \label{rm:equ}
 The view $V$ and simulator $S$ are computationally  indistinguishable $V \stackrel{c}{\equiv} S$, if they have the same list of values or have values that are generated according to the same distribution and independent from other parameters \citep{conf:cloutdquadratic}.
 \end{remark}
 We denote the quantities obtained by the simulator by $\widetilde{()}$, which follow the same distribution but are otherwise different from the quantities of the views. We may omit the time index $k$ from the views and simulators for simplicity. Here, the coins are random numbers that are used for the encryption process and key generation. Further information that is exchanged between other parties over an encrypted channel is denoted by $\enc{\Gamma_X}$ for coalition $X$. Note that the encrypted channel uses extra keys different from the homomorphic encryption keys and uses double encryption to protect privacy.
 
\subsection{Proof of Theorem \ref{th:alg1pricol}}
\label{prf:alg1}

 \begin{proof}[\unskip\nopunct]
The proof consists of three types of coalitions as described bellow.

 \subsubsection{Coalition of sensors $s$} \label{sec:ssensorzono}
The strip information is considered as the input to the sensor and appears as part of the view and the simulator of the coalition. We denote the view of coalition $s$ consisting of the set of sensors $s=\{s_{1},\dots,s_{t}\}$ by $V^{\Pi}_{s}$, which is defined as the combination of every sensor view and given by
\begin{align}
    V^{\Pi}_{s} =\big(&V^{\Pi}_{s_{1}},\dots,V^{\Pi}_{s_{t}} \big) \nonumber\\
    =  \big(& H_{s,k},y_{s,k},r_{s,k},\enc{y_{s,k}},\text{coins}_{s},\textit{pk}, \enc{\Gamma_s}
    \big), 
    \label{eq:viewsencol}
\end{align}
where the subscript $s$ on $H_{s,k},y_{s,k},r_{s,k}$ denotes an array of strip information of the coalition. 
The sensors only submit their encrypted data to the
aggregator. Hence, a simulator, denoted by $S_s$, consists of the input and output and by generating 
$\widetilde{\enc{\Gamma_s}}, \widetilde{\enc{y_{s,k}}}$ and  $\widetilde{\text{coins}}_{s}$, i.e.
\begin{align}
    S_{s} =\big(&\textit{pk},H_{s,k},y_{s,k},r_{s,k},\widetilde{\enc{y_{s,k}}},\widetilde{\text{coins}}_{s}, \widetilde{\enc{\Gamma_s}}
    \big).
\end{align}
The $\widetilde{\text{coins}}_{s}$ are generated according to the same distribution of $\text{coins}_{s}$ and are independent from other parameters, where the same is true for $\widetilde{\enc{\Gamma_s}}$ and $\enc{\Gamma_s}$ as well as $\widetilde{\enc{y_{s,k}}}$ and $\enc{y_{s,k}}$. Therefore, we conclude that $S_{s} \stackrel{c}{\equiv} V^{\Pi}_{s}$.
   

Moreover, the information contained in each strip is independent from all others. Thus, the coalition strips cannot be used to infer new information about other strips. The information in each iteration is different from other iterations. That is why we considered only a single step in the previous proof. In the following two subsections, we will prove that the view of each coalition after $K\in\mathbb{N}^+$ iterations of the protocol is computationally indistinguishable from the view of a simulator that executes $K$ iterations.
\subsubsection{Coalition of  sensors $s$  and aggregator}
\label{sec:ssenaggzono}

The view of the aggregator is denoted by  $V^{\Pi}_{a}$. We denote the view of a coalition consisting of a set of sensors by $s=\{s_{1},\dots,s_{t}\}$ and the aggregator by $V^{\Pi}_{sa}$ which is defined by
\begin{align}
    V^{\Pi}_{sa} =\big(& V^{\Pi}_{s},V^{\Pi}_{a}  \big)= \big( V^{\Pi,K}_{s},V^{\Pi,K}_{a}  \big),
    \label{eq:vsapr1}
\end{align}
where $V^{\Pi,K}_{s}$ and $V^{\Pi,K}_{a}$ are the views of the aggregator and coalition of sensors, after executing $K$ iterations, respectively, and are given by
\begin{align}\label{eq_add}
 V^{\Pi,k+1}_{s}=( V^{\Pi,k}_{s},I_{s}^{k+1}), \quad  V^{\Pi,k+1}_{a}=( V^{\Pi,k}_{a},I_{a}^{k+1})
\end{align}
$\forall k=0, 1,\dots, K-1$, where $I_{s}^k$ and $I_{a}^k$ are the newly added data points at the $k$-th iteration for the coalition of sensors $V^{\Pi,0}_{s}=I_{s}^0$ and aggregator $V^{\Pi,0}_{a}=I_{a}^0$. The view of the aggregator contains encrypted strips $\zono{H_{s,k},\enc{y_{s,k}},r_{s,k}}$ from the sensors, initial set $\langle \enc{\hat{c}_{q,0}},\hat{G}_{q,0} \rangle$ from the query node, and the estimated set $\langle \enc{\hat{c}_{a,k}}, \hat{G}_{a,k}\rangle$ at $k$-th iteration. Let us denote the strip information of the sensors at the $k$-th iteration, which are not part of the coalition by subscript $r$, i.e., $\zono{H_{r,k},\enc{y_{r,k}},r_{r,k}}$, $k=0, 1,\dots, K-1$. Then, $I_{a}^k$ and $I_{s}^k$ are 
\begin{align}
I_{a}^k  =\big(& H_{s,k},\enc{y_{s,k}},r_{s,k} , H_{r,k},\enc{y_{r,k}},r_{r,k}, \enc{\hat{c}_{q,0}}, \nonumber\\ 
 &\hat{G}_{q,0}, \enc{\hat{c}_{a,k}},  \hat{G}_{a,k}, \text{coins}_{a}, \textit{pk},q, F, Q_k
    \big),\label{eq:viewaggcol}\\
    I_{s}^k  =\big(& H_{s,k},y_{s,k},r_{s,k},\enc{y_{s,k}},\text{coins}_{s},\textit{pk}, \enc{\Gamma_s}
    \big),
    \label{eq:viewsencol2}
\end{align}
%
%
where $\mathcal{Z}_{q,0} = \zono{\hat{c}_{q,0}, \hat{G}_{q,0}}$ is the initial zonotope at the query node and $\mathcal{Z}_{a,k}=\zono{\hat{c}_{a,k},  \hat{G}_{a,k}}$ is the estimated zonotope on the aggregator side at time step $k$. The view of the coalition $V^{\Pi}_{sa}$ is constructed from \eqref{eq:vsapr1}--\eqref{eq:viewsencol2}. Let the simulator of the coalition be denoted by $S_{sa}=S_{sa}^{K}$, where $S_{sa}^{K}$ is the simulator after executing $K$ iterations. The simulator $S_{sa}$ can be iteratively constructed by combining the values obtained at each time step $k$ as follows:
 \begin{align}
S_{sa}^{k+1}=(S^{k}_{sa},I^{S,k+1}_{sa} ),\quad k=0, 1,\dots, K-1,
 \end{align}
 where $I^{S,k+1}$ is the portion of the simulator generated at iteration $k+1$, which is given by
\begin{align*}
I^{S,k}_{sa}=\big(& H_{s,k},\widetilde{\enc{y_{s,k}}},r_{s,k} , H_{r,k},\widetilde{\enc{y_{r,k}}},r_{r,k}, \widetilde{\enc{\hat{c}_{q,0}}}, \hat{G}_{q,0},\nonumber\\ 
 & \widetilde{\enc{\hat{c}_{a,k}}},  \hat{G}_{a,k}, \widetilde{\text{coins}}_{sa}, q, F, Q_k,  y_{s,k},\textit{pk}, \widetilde{\enc{\Gamma_s}}
\big)
\end{align*}
and where the values are computed or generated as follows:
\begin{enumerate}
    \item Generate $\widetilde{\enc{\Gamma_s}},\widetilde{\enc{\bar{c}_{q,0}}}$,$\widetilde{\enc{\hat{c}_{a,k}}}$, $\widetilde{\enc{y_{r,k}}}$, and $\widetilde{\enc{y_{s,k}}}$ according to the same distribution of $\enc{\Gamma_s}$, $\enc{\bar{c}_{q,0}}$, $\enc{\hat{c}_{a,k}}$, $\enc{y_{r,k}}$ and $\enc{y_{s,k}}$, respectively.
    \item Compute $\hat{G}_{a,k}$ according to \eqref{eq:G_lambda}.
    \item Let the combination of all coins of the parties be $\text{coins}_{sa} = (\text{coins}_{a},\text{coins}_{s})$. Generate $\widetilde{\text{coins}}_{sa}$ according to the distribution of $\text{coins}_{sa}$.
\end{enumerate}

Based on this generation scheme, the values $\widetilde{\enc{}}$ and $\enc{}$ are indistinguishable and all remaining variables in $I^{S,k+1}_{sa}$ are either public or feasible through the protocol steps. After all iteration steps, we end up with a simulator that satisfies $S_{sa} \stackrel{c}{\equiv} V^{\Pi}_{sa}.$

The second part of the proof is about inferring extra private information from the input and output. The coalition's target is to determine the private measurement of the remaining sensors $y_{r,k}$. Note that the relation between $\enc{y_{s,k}}$ and $\enc{y_{r,k}}$ is characterized by \eqref{eq:undercolsa}.
\begin{align}
 \sum\limits_{j\in\mathcal{N}_{r}} \lambda_{j,k} \enc{y_{j,k}}  =& \sum\limits_{j\in\mathcal{N}} ( \lambda_{j,k} H_{j,k} - 1) \enc{\hat{c}_{a,k-1}} \oplus \enc{\bar{c}_{a,k}}  \nonumber\\ 
 &   \ominus \underbrace{\sum\limits_{j\in\mathcal{N}/r} \lambda_{j,k} \enc{y_{j,k}}}_{\text{known to the coalition in plaintext}}, \label{eq:undercolsa}
\end{align}
where $\mathcal{N}_{r}$ is the set of the remaining sensors. Since the coalition does not have the private key and the query node sends the initial encrypted center $\enc{\hat{c}_{a,0}}$, we end up with an underdetermined system in \eqref{eq:undercolsa}. 

\subsubsection{Coalition of  sensors $s$  and query node}\label{sec:ssenqueryzono}
We denote the view of a coalition consisting of a set of sensors by $s=\{s_{1},\dots,s_{t}\}$ and define the query as
%
\begin{align}
V^{\Pi}_{sq} =\big(& V^{\Pi}_{s},V^{\Pi}_{q}  \big)= \big( V^{\Pi,K}_{s},V^{\Pi,K}_{q}  \big),
\end{align}
where 
\begin{align}\label{eq_add2}
V^{\Pi,k+1}_{s}=( V^{\Pi,k}_{s},I_{s}^{k+1}), \quad  V^{\Pi,k+1}_{q}=( V^{\Pi,k}_{q},I_{q}^{k+1}),
\end{align}
$\forall k=0, 1,\dots, K-1,$ where $I_{s}^k$ is given in \eqref{eq:viewsencol2}, and $I_{q}^k$ are the newly added data points from the $k$-th iteration for the query node  with $ V^{\Pi,0}_{q}=I_{q}^0$ such that
\begin{align}\label{eq_add3}
I_{q}^k=\big(\hat{c}_{q,0},\hat{G}_{q,0},\hat{c}_{a,k},\hat{G}_{a,k},\enc{\hat{c}_{a,k}}, \text{coins}_{q}, \enc{\Gamma_{sq}},\textit{pk},\textit{sk}\big).
\end{align}

The view of the coalition $V^{\Pi}_{sq}$ is constructed from \eqref{eq:viewsencol2}, \eqref{eq_add2} and \eqref{eq_add3}. The construction of the simulator is similar to Section \ref{sec:ssenaggzono}. Thus, we focus on the values added in the $k$-th iteration to the simulator. Let the combination of all coins of the parties be denoted by $\text{coins}_{sq} = (\text{coins}_{q},\text{coins}_{s})$. The inputs and outputs to the coalition are \big($\textit{pk},\textit{sk},H_{s,k},y_{s,k},r_{s,k},\hat{c}_{a,k},\hat{G}_{a,k},\hat{c}_{q,0},\hat{G}_{q,0}$\big). Thus, the simulator $S_{sq}^k$ can be easily generated by
\begin{align}
    S_{sq}^k =\big(&\textit{pk},\textit{sk},H_{s,k},y_{s,k},r_{s,k},\hat{c}_{q,0},\hat{G}_{q,0},\hat{c}_{a,k},\hat{G}_{a,k},\widetilde{\enc{\hat{c}_{a,k}}},\nonumber\\
    &\widetilde{\text{coins}}_{sq},\widetilde{\enc{\Gamma_{sq}}}, S_{sq}^{k-1} \big).
\end{align}

The tuples $\left(\enc{\hat{c}_{a,k}}, \text{coins}_{sq}, \enc{\Gamma_{sq}}\right)$ and $\left(\widetilde{\enc{\hat{c}_{a,k}}}, \widetilde{\text{coins}}_{sq}, \widetilde{\enc{\Gamma_{sq}}}\right)$ are generated according to the same distribution and are independent from other parameters. Therefore, $S_{sq}^k \stackrel{c}{\equiv} (I_{s}^k,I_{q}^k),$ which leads to 
\begin{equation}
   S_{sq} \stackrel{c}{\equiv} V^{\Pi}_{sq}.
\end{equation}

In case of a coalition between $s$ sensors and the query node, the aim would be to find the measurements of the remaining group, denoted by ${N}_{r}$ with size $m_r$. Rewriting \eqref{eq:undercolsa} after decryption -- as the query has the Paillier private key \textit{sk} -- results in
 \begin{align}\label{eq_privacy}
\Lambda_{r,k}Y_{r,k}=z_{s,k},
 \end{align}
where
\begin{align*}
z_{s,k}&=\sum\limits_{j\in\mathcal{N}} ( \lambda_{j,k} H_{j,k}) \hat{c}_{a,k-1} + \bar{c}_{a,k} - \!\! \sum\limits_{j\in\mathcal{N}/r} \lambda_{j,k} y_{j,k}, \\
\Lambda_{r,k}&=[\lambda_{j_1,k},\lambda_{j_2,k},\dots,\lambda_{j_{m_r},k}]\in\mathbb{R}^{n\times pm_r},\\
Y_{r,k}&=[y_{j_1,k}^T,y_{j_2,k}^T,\dots,y_{j_{m_r},k}^T]^T\in\mathbb{R}^{ pm_r},
\end{align*}
where $z_{s,k}$ is known to the coalition given that $\lambda_{j,k}$ is computed based on the generator matrix. To find the conditions at which the privacy of $Y_{r,k}$ is ensured, we show that there is no unique retrieval for $Y_{r,k}$. This non-unique retrieval requires that \eqref{eq_privacy} has multiple solutions. According to \citep[Theorem 6.4]{schott2016matrix}, $\tilde{Y}_{r,k}$ is a solution of \eqref{eq_privacy} for any $X_r \in\mathbb{R}^{pm_r}$ with
 \begin{align}\label{eq_privacy2}
\tilde Y_{r,k}=\Lambda_{r,k}^-z_{s,k}+(I_{pm_r}-\Lambda_{r,k}^-\Lambda_{r,k})X_r,
 \end{align}
where $\Lambda_{r,k}^-$ is any generalized inverse of $\Lambda_{r,k}$. For every solution $\tilde{Y}_{r,k}$ of \eqref{eq_privacy} there is an $X_r$. For $I_{pm_r}-\Lambda_{r,k}^-\Lambda_{r,k}= 0$, the system is consistent and thus has one solution \citep[Theorem 6.1]{schott2016matrix}. Therefore, we aim to find conditions at which $I_{pm_r}-\Lambda_{r,k}^-\Lambda_{r,k}\neq 0$ to ensure privacy. We have $rank(\Lambda_{r,k}^-\Lambda_{r,k})\leq \min\{pm_r,n\}$ according to \citep[Theorem 2.8]{schott2016matrix}. Thus, under the condition $pm_r>n$, we have $I_{pm_r}-\Lambda_{r,k}^-\Lambda_{r,k}\neq 0$ which ensures the privacy of $Y_{r,k}$.
\end{proof}


\subsection{Proof of Theorem \ref{th:alg3pricol}}
\label{prf:alg3}

\begin{proof}[\unskip\nopunct]
In the following proof, we   consider the view and simulation for one step (i.e.,  $k$-th step) for notational convenience. The proof for $K\in\mathbb{N}^+$ steps is similar to the proof of Theorem  \ref{th:alg1pricol}. We are going to prove the privacy against the three coalitions as follows:

 \subsubsection{Coalition of  sensors $s$}
The strips information is considered to be an input to the sensor and appears as part of the view and the simulator of the coalition. The strips information is exactly the same as for zonotopes. Thus, the proof is similar to section \ref{sec:ssensorzono} and is therefore omitted.

\subsubsection{Coalition of  sensors $s$  and aggregator}

The aggregator has encrypted strips $(H_{s,k},\enc{y_{s,k}}, R_{s,k},H_{r,k},\\ \enc{y_{r,k}},R_{r,k})$ from the sensors, the initial constrained zonotope $\zono{\enc{\hat{c}_{q,0}}, \hat{G}_{q,0}, \enc{\hat{b}_{q,0}}, \hat{A}_{q,0}}$ from the query node, and estimated constrained zonotope $\zono{\enc{\hat{c}_{g,k}}, \hat{G}_{g,k}, \enc{\hat{b}_{g,k}}, \hat{A}_{g,k}}$ at each $k$-iteration. The view of the coalition is defined as
\begin{align}
    V^{\Pi}_{sa} =\big(& V^{\Pi}_{s},V^{\Pi}_{a}  \big) \nonumber\\
               =\big(& V^{\Pi}_{s}, H_{s,k},\enc{y_{s,k}},R_{s,k} , H_{r,k},\enc{y_{r,k}},R_{r,k},\enc{\hat{c}_{q,0}} ,\hat{G}_{q,0},\enc{\hat{b}_{q,0}},\nonumber\\
    & \hat{A}_{q,0},\enc{\hat{c}_{g,k}} ,\hat{G}_{g,k},\enc{\hat{b}_{g,k}}, 
                \hat{A}_{g,k}, \text{coins}_{sa},\textit{pk}, q, F,Q_k \big)\nonumber\\
                \stackrel{\eqref{eq:viewsencol}}{=} \big(& H_{s,k},y_{s,k},R_{s,k},\enc{y_{s,k}}, 
                  H_{r,k},\enc{y_{r,k}},R_{r,k} ,\enc{\hat{c}_{q,0}} ,\hat{G}_{q,0}, \enc{\hat{b}_{q,0}} ,\nonumber\\
    & \hat{A}_{q,0},\enc{\hat{c}_{g,k}},\hat{G}_{g,k},\enc{\hat{b}_{g,k}} ,\hat{A}_{g,k},
            \text{coins}_{sa},\textit{pk},q, F,Q_k \big).
\end{align}

The simulation is the same as in Section \ref{sec:ssenaggzono}, except for the additional information contained in a constrained zonotope, i.e., the constraints. This results in
\begin{align}
    S_{sa} =\big(& H_{s,k},y_{s,k},R_{s,k},\widetilde{\enc{y_{s,k}}}, 
                  H_{r,k},\widetilde{\enc{y_{r,k}}},R_{r,k} ,\widetilde{\enc{\hat{c}_{q,0}}} ,\hat{G}_{q,0}, \nonumber\\
    &\widetilde{\enc{\hat{b}_{q,0}}} ,\hat{A}_{q,0},\widetilde{\enc{\hat{c}_{g,k}}} ,\hat{G}_{g,k},\widetilde{\enc{\hat{b}_{g,k}}} ,\hat{A}_{g,k} ,\widetilde{\text{coins}}_{sa},\textit{pk},q, F\nonumber\\
    &,Q_k \big).
\end{align}
We arrive at a simulator that satisfies $S_{sa} \stackrel{c}{\equiv} V^{\Pi}_{sa}$. Thus, similarly to Section \ref{sec:ssenaggzono}, the coalition cannot infer extra information from the input and the output.

\subsubsection{Coalition of  sensors $s$  and query node}

The view of the coalition consists of the view of the sensors $V^{\Pi}_{s}$ and the view of the query node $V^{\Pi}_{q}$ which consist of the initial estimated constrained zonotope $\zono{\hat{c}_{q,0},\hat{G}_{q,0},\hat{A}_{q,0},\hat{b}_{q,0}}$ and resultant estimated set $\zono{\hat{c}_{a,k},\hat{G}_{a,k},\hat{A}_{a,k},\hat{b}_{a,k}}$ at each $k$-iteration as follows:
\begin{align}
    V^{\Pi}_{sq} =\big(& V^{\Pi}_{s},V^{\Pi}_{q}  \big)\nonumber\\
    =\big(& V^{\Pi}_{s},\hat{c}_{q,0},\hat{G}_{q,0},\hat{A}_{q,0},\hat{b}_{q,0},\hat{c}_{a,k},\hat{G}_{a,k},\hat{A}_{a,k},\hat{b}_{a,k},\text{coins}_{s},\enc{\Gamma_{s}},
    \nonumber\\
    &\textit{pk},\textit{sk}   \big) \nonumber\\
    \stackrel{\eqref{eq:viewsencol}}{=}& \big( H_{s,k},y_{s,k},R_{s,k},\enc{y_{s,k}}, 
                  H_{r,k},\enc{y_{r,k}},R_{r,k} ,\hat{c}_{q,0},\hat{G}_{q,0},\hat{A}_{q,0},\hat{b}_{q,0},\nonumber\\
    & \hat{c}_{a,k},\hat{G}_{a,k},\hat{A}_{a,k},\hat{b}_{a,k}, \text{coins}_{sq}, \enc{\Gamma_{sq}},\textit{pk},\textit{sk}  \big).
\end{align}

The simulator will be again similar to Section \ref{sec:ssenqueryzono} after adding and generating the constrained zonotope information, specifically
\begin{align}
    S_{sq} =\big(& H_{s,k},y_{s,k},R_{s,k},\widetilde{\enc{y_{s,k}}}, 
                  H_{r,k},\widetilde{\enc{y_{r,k}}},R_{r,k} ,\hat{c}_{q,0},\hat{G}_{q,0}, \hat{A}_{q,0},\hat{b}_{q,0},\nonumber\\
    &\hat{c}_{a,k},\hat{G}_{a,k},\hat{A}_{a,k},\hat{b}_{a,k}, \widetilde{\text{coins}}_{sq}, \widetilde{\enc{\Gamma_{sq}}},\textit{pk},\textit{sk}  \big).
\end{align}

As before, the generated values $\widetilde{\enc{y_{s,k}}}, \widetilde{\enc{y_{r,k}}},\widetilde{\text{coins}}_{sq}$ and $\widetilde{\enc{\Gamma_{sq}}}$ are generated according to the distribution of the original values and are independent from other parameters. Therefore $S_{sq} \stackrel{c}{\equiv} V^{\Pi}_{sq}$.

The coalition aims to find the measurements of the remaining group, denoted by ${N}_{r}$ with size $m_r$. Note  that $\Lambda_{a,k}$ is chosen by the aggregator and not known to the query; additionally, it is also chosen at random and not dependant on publicly shared generator matrix. Thus, computing the measurement $y_{r,k}$ according to \eqref{eq_privacy} is not valid anymore. In contrast to Theorem~\ref{th:alg1pricol}, privacy can be guaranteed in all cases. 
\end{proof} 
\subsection{Proof of Theorem \ref{th:alg2pricol}}
\label{prf:alg2}

In the following proof, we consider the view and simulation for one step (i.e.,  $k$-th step) for notational convenience. The proof for $K\in\mathbb{N}^+$ steps is similar to the proof of Theorem  \ref{th:alg1pricol}. We prove again the privacy against the following three coalitions:

 \subsubsection{Coalition of sensors groups $g$:}

We define the view of a coalition consisting of a of set of sensor groups $g=\{g_{1},\dots,g_{t}\}$ by $V^{\Pi}_{g}$ by
\begin{align}
    V^{\Pi}_{g} &= \big(V^{\Pi}_{g_{1}},\dots,V^{\Pi}_{g_{t}} \big)\nonumber \\
    &=  \big(\textit{pk}, H_{g,k},y_{g,k},R_{g,k},\bar{G}_{g,k},\bar{c}_{g,k},\enc{\bar{c}_{g,k}}, \enc{\Gamma_g},coin_g
    \big),
\end{align}
where the subscript $g$ denotes the variables owned by the coalition. The sensor groups only submit their encrypted data to the aggregator. Hence, a simulator $S_g$, defined by
\begin{align}
    S_{g} =& \big(\textit{pk}, H_{g,k}, y_{g,k}, R_{g,k}, \bar{G}_{g,k}, \bar{c}_{g,k}, \widetilde{\enc{\bar{c}_{g,k}}}, \widetilde{\enc{\Gamma_g}},\widetilde{coin}_g \big),
\end{align}
is obtained by generating $\widetilde{\enc{\bar{c}_{g,k}}}, \widetilde{\enc{\Gamma_g}}$ and $\widetilde{\text{coins}}_{g}$ according to the distribution of ($\enc{\bar{c}_{g,k}}, \enc{\Gamma_g},coin_g$) and are independent from other parameters. Therefore, we conclude that $ S_{g} \stackrel{c}{\equiv} V^{\Pi}_{g}$. 

Moreover, the resulting zonotopes from the sensor groups are independent. As a result, the coalition zonotopes cannot be used to infer new information about other zonotopes. 

 \subsubsection{Coalition of sensor groups  $g$  and the aggregator:} \label{sec:gsengrpaggzono}

The view of the coalition is defined by
\begin{align}
    V^{\Pi}_{ga} =& \big( V^{\Pi}_{g},V^{\Pi}_{a}  \big)
\end{align}
with
\begin{align}
    V^{\Pi}_{a} =&  \big(\enc{\bar{c}_{g,k}} ,\bar{G}_{g,k},  \enc{\bar{c}_{r,k}} ,\bar{G}_{r,k},\enc{\hat{c}_{a,k}}, \hat{G}_{a,k}, q, F, Q_k,\text{coins}_a,\textit{pk}\big)
\end{align}
where $\enc{\bar{c}_{r,k}}$ and $\bar{G}_{r,k}$ represents the encrypted center and the generators of the remaining sensor groups which are not part of the coalition. The simulator, denoted by $S_{ga}$, can be constructed from the input and output ($H_{g,k}, R_{g,k}, F, \textit{pk}, q,Q_k,y_{g,k}$). Specifically:

\begin{enumerate}
    \item Add $H_{r,k}$ and $R_{r,k}$ as they are public information. 
    \item Compute $\bar{G}_{g,k}$ and $\bar{G}_{r,k}$ according to \eqref{eq:G_lambda}.
    \item Compute $\hat{G}_{a,k}$ according to \eqref{eq:Gtildehatzonotimeupdate}.
    \item Generate $\widetilde{\enc{\bar{c}_{g,k}}}, \widetilde{\enc{\bar{c}_{r,k}}}, \widetilde{\enc{\Gamma_g}}$, and $\widetilde{\enc{\hat{c}_{a,k}}}$ according to the distributions of the original values. 
    \item Let the combination of coins of all parties be $\text{coins}_{ga} = (\text{coins}_{a},\text{coins}_{g})$. Generate $\widetilde{\text{coins}}_{ga}$ according to the distribution of $\text{coins}_{ga}$.
    \item Compute $\bar{c}_{g,k}$ according to \eqref{eq:C_lambda}.
\end{enumerate}
We end up with the simulator
\begin{align}
    S_{ga} =  \big(&\textit{pk}, H_{r,k}, R_{r,k}, H_{g,k}, y_{g,k}, R_{g,k}, \bar{G}_{g,k}, \bar{c}_{g,k}, \widetilde{\enc{\bar{c}_{g,k}}},\widetilde{\enc{\bar{c}_{r,k}}},  \widetilde{\enc{\Gamma_g}},\nonumber \\ 
    &\bar{G}_{r,k},\widetilde{\enc{\hat{c}_{a,k}}},\hat{G}_{a,k}, q, F, Q_k, \widetilde{\text{coins}}_{ga},\textit{pk}\big).
\end{align}

Thus, we find that $ S_{ga} \stackrel{c}{\equiv} V^{\Pi}_{ga}$. The target of this coalition is to get the zonotopes of the remaining groups, denoted by ${N}_{r}$ with size $d_r$. The centers of the zonotopes are related by
\begin{align}
    \sum\limits_{j\in\mathcal{N}_{r}}w^{j}_{a,k} \enc{\bar{c}_{g_j,k}} \oplus \enc{\grave{c}_{a,k}} \sum\limits_{j\in\mathcal{N}}w^{j}_{a,k} &= \sum\limits_{j\in\mathcal{N}/r}w^{j}_{a,k} \enc{\bar{c}_{g_j,k}}.  \label{eq:cenga}
\end{align}

The right hand side of \eqref{eq:cenga} is known to the coalition. 
However, since the coalition does not have the private key, the privacy of the centers of the remaining group can be guaranteed.



\subsubsection{Coalition of sensor groups  $g$  and the query:}

The view of the coalition is defined as $V^{\Pi}_{gq}$ where $V^{\Pi}_{gq} = \big( V^{\Pi}_{g},V^{\Pi}_{q}  \big)$ with $V^{\Pi}_{q}$ given by
\begin{align}
    V^{\Pi}_{q} =&  \big( \enc{\hat{c}_{a,k}},\hat{c}_{a,k}, \hat{G}_{a,k}, q, F, Q_k,
 \text{coins}_q,\textit{pk},\textit{sk},\enc{\Gamma_q} \big).
\end{align}

Constructing the simulator $S_{gq}$ from the inputs and outputs of the coalition as done before results in 
\begin{align}
S_{gq} = \big(&\textit{pk}, \textit{sk},H_{g,k}, y_{g,k}, R_{g,k}, \bar{G}_{g,k}, \bar{c}_{g,k}, \widetilde{\enc{\Gamma_q}} \widetilde{\enc{\hat{c}_{a,k}}},\hat{c}_{a,k}, \hat{G}_{a,k},\nonumber \\
&\widetilde{\text{coins}}_{gq}\big),
\end{align}
which implies that $S_{gq} \stackrel{c}{\equiv} V^{\Pi}_{gq}$. The target of this coalition is to get the zonotopes of the remaining group, denoted as before by $\mathcal{N}_{r}$ with size $d_r$. Rewriting \eqref{eq:cenga} after decryption -- as the query has the Paillier private key \textit{sk} -- results in
 \begin{align}\label{eq:wprivacy}
W_{r,k}C_{r,k}=z_{g,k},
 \end{align}
with
\begin{align}
z_{g,k}&=\sum\limits_{j\in\mathcal{N}/r}w^{j}_{a,k} \bar{c}_{g_j,k}  - \grave{c}_{a,k} \sum\limits_{j\in\mathcal{N}}w^{j}_{a,k}, \\
W_{r,k}&=[w_{a,k}^{j_1}I_n,w_{a,k}^{j_2}I_n,\dots,w_{a,k}^{j_{d_r}}I_n]\in\mathbb{R}^{n\times nd_r},\\
	C_{r,k}&=[c_{j_1,k}^T,c_{j_2,k}^T,\dots,c_{j_{d_r},k}^T]^T\in\mathbb{R}^{ nd_r},
\end{align}
where $z_{g,k}$ is known to the coalition. Similarly to the proof of Theorem \ref{th:alg1pricol} and according to \citep[Theorem 6.4]{schott2016matrix}, $\tilde{C}_{r,k}$ is a solution of \eqref{eq:wprivacy} for any $X_r \in\mathbb{R}^{nd_r}$ where
 \begin{align}\label{eq_privacy2}
\tilde C_{r,k}=W_{r,k}^-z_{g,k}+(I_{nd_r}-W_{r,k}^-W_{r,k})X_r,
 \end{align}
and where $W_{r,k}^-$ is any generalized inverse of $W_{r,k}$. For every solution $\tilde{C}_{r,k}$ of \eqref{eq_privacy} there is a $X_r$. If $I_{nd_r}-W_{r,k}^-W_{r,k}= 0$, we end up with a consistent system with one solution \citep[Theorem 6.1]{schott2016matrix}. Thus, we aim to find conditions at which $I_{nd_r}-W_{r,k}^-W_{r,k}\neq 0$ to ensure privacy. We have $rank(W_{r,k}^-W_{r,k})\leq \min\{nd_r,n\}$ according to \citep[Theorem 2.8]{schott2016matrix}. Thus, under the condition $d_r>1$, it follows that $I_{pd_r}-W_{r,k}^-W_{r,k}\neq 0$ which ensures privacy of $C_{r,k}$.
\vspace{-2mm}
\subsection{Proof of Theorem \ref{th:alg4pricol}}
\label{prf:alg4}

In the following proof, we consider the view and simulation for one step (i.e.,  $k$-th step) for notational convenience. The proof for $K\in\mathbb{N}^+$ steps is similar to the proof of Theorem  \ref{th:alg1pricol}. We are going to prove the privacy against the three coalitions as follows:

 \subsubsection{Coalition of sensor groups  $g$ :}
The view of the coalition can be defined by
\begin{align}
    V^{\Pi}_{g} = \big(&V^{\Pi}_{g_{1}},\dots,V^{\Pi}_{g_{t}} \big)
    =  \big(\textit{pk}, H_{g,k},y_{g,k},R_{g,k},\bar{G}_{g,k},\bar{c}_{g,k},\enc{\bar{c}_{g,k}},\nonumber \\
    &\bar{A}_{g,k}, \bar{b}_{g,k},\enc{\bar{b}_{g,k}}, \enc{\Gamma_g},coin_g
    \big).
    \label{eq:vsengrp}
\end{align}
Again, sensors only submit their encrypted data to the
aggregator. Hence, a simulator $S_s$ given by
\begin{align}
    S_{g} = \big(&\textit{pk}, H_{g,k}, y_{g,k}, R_{g,k}, \bar{G}_{g,k}, \bar{c}_{g,k}, \widetilde{\enc{\bar{c}_{g,k}}},\bar{A}_{g,k}, \bar{b}_{g,k},\widetilde{\enc{\bar{b}_{g,k}}},   \widetilde{\enc{\Gamma_g}},\nonumber \\
    &\widetilde{coin}_g \big),
\end{align}
is obtained by generating 
$\widetilde{\enc{\bar{c}_{g,k}}}, \widetilde{\enc{\Gamma_g}},\widetilde{\enc{\bar{b}_{g,k}}}$ and $\widetilde{\text{coins}}_{g}$. 
The generated and the original values are generated according to the same distribution and are independent from other parameters. Therefore, we conclude that $S_{g} \stackrel{c}{\equiv} V^{\Pi}_{g}$.
  

Moreover, the resulting constrained zonotopes from the sensor groups are independent. Thus, the coalition zonotopes cannot be used to infer new information about other zonotopes. 

 \subsubsection{Coalition of sensor groups  $g$  and the aggregator:}

The view of the coalition, denoted by $V^{\Pi}_{ga}$, is
\begin{align}
    V^{\Pi}_{ga} = \big(& V^{\Pi}_{g},V^{\Pi}_{a}  \big) \nonumber\\
                 =\big(& V^{\Pi}_{g},\enc{\bar{c}_{g,k}} ,\bar{G}_{g,k},\bar{A}_{g,k}, \enc{\bar{b}_{g,k}} ,\enc{\bar{c}_{r,k}} ,\bar{G}_{r,k}, \bar{A}_{r,k},\enc{\bar{b}_{r,k}},\enc{\hat{c}_{a,k}},\nonumber \\
    & \hat{G}_{a,k},\hat{A}_{a,k},\enc{\hat{b}_{a,k}}, q, F, Q_k, \text{coins}_a,\textit{pk} \big)  \nonumber\\
    \stackrel{\eqref{eq:vsengrp}}{=}& \big(H_{r,k},R_{r,k}, H_{g,k},y_{g,k},R_{g,k},\bar{c}_{g,k},\bar{G}_{g,k}, \enc{\bar{c}_{g,k}}, \bar{A}_{g,k}, \bar{b}_{g,k},    \enc{\bar{b}_{g,k}},\nonumber \\
    &\enc{\bar{c}_{r,k}},\bar{G}_{r,k}, \bar{A}_{r,k},\enc{\bar{b}_{r,k}},
                 \enc{\hat{c}_{a,k}}, \hat{G}_{a,k},  \hat{A}_{a,k},\enc{\hat{b}_{a,k}},q, F, Q_k,\nonumber \\
    &\text{coins}_{ga},\textit{pk} \big)  
\end{align}

where $ \langle \enc{\bar{c}_{r,k}},\bar{G}_{r,k},\bar{A}_{r,k},\enc{\bar{b}_{r,k}}  \rangle$, represents the encrypted constrained zonotopes of the sensor groups which are not part of the coalition. The simulator, denoted by $S_{ga}$, can be constructed given the input and output ($H_{g,k}, R_{g,k}, F, \textit{pk}, q, Q_k, y_{g,k},\bar{b}_{g,k}, \bar{c}_{g,k}$) as follows:
\begin{enumerate}
    \item Add $H_{r,k}$, $R_{r,k}$ as they are public information. 
    \item Compute $\bar{G}_{g,k}$ and $\bar{G}_{r,k}$ according to \eqref{eq:G_lambdapr3}.
    \item Compute $\bar{A}_{g,k}$ and $\bar{A}_{r,k}$ according to \eqref{eq:Abarconstzono}.
    \item Compute $\hat{G}_{a,k}$ according to \eqref{eq:gdiff} and reduction operation similar to \eqref{eq:GAhatpr3}.
    \item Compute $\hat{A}_{a,k}$ according to \eqref{eq:Abgravpr4} and reduction operation similar to \eqref{eq:GAhatpr3}.
    \item Generate $\widetilde{\enc{\bar{c}_{g,k}}}, \widetilde{\enc{\bar{c}_{r,k}}}$, and $\widetilde{\enc{\hat{c}_{a,k}}}$ according to the distributions of the original values.
    \item Generate $\widetilde{\enc{\bar{b}_{g,k}}}, \widetilde{\enc{\bar{b}_{r,k}}}$, and $\widetilde{\enc{\hat{b}_{a,k}}}$ according to the distributions of the original values.
    \item Let the combination of the coins of all parties be $\text{coins}_{ga} = (\text{coins}_{a},\text{coins}_{g_1},\dots,\text{coins}_{g_t})$. Generate $\widetilde{\text{coins}}_{ga}$ according to the distribution and of $\text{coins}_{ga}$.
\end{enumerate}
We end up with the following simulator
\begin{align}
    S_{ga} =  \big(&H_{r,k},R_{r,k}, H_{g,k},y_{g,k},R_{g,k},\bar{c}_{g,k},\bar{G}_{g,k}, \widetilde{\enc{\bar{c}_{g,k}}}, \bar{A}_{g,k}, \bar{b}_{g,k},    \widetilde{\enc{\bar{b}_{g,k}}},\nonumber \\
    & \widetilde{\enc{\bar{c}_{r,k}}},\bar{G}_{r,k}, \bar{A}_{r,k}, \widetilde{\enc{\bar{b}_{r,k}}}, \widetilde{\enc{\hat{c}_{a,k}}}, \hat{G}_{a,k},\hat{A}_{a,k},    \widetilde{\enc{\hat{b}_{a,k}}},q, F,Q_k,\nonumber \\
    &\widetilde{\text{coins}}_{ga},\textit{pk} \big).
\end{align}

Thus, we find that $S_{ga} \stackrel{c}{\equiv} V^{\Pi}_{ga}$. Similarly to Section \ref{sec:gsengrpaggzono}, the coalition is not be able to infer information about the constrained zonotopes of the remaining group.

\subsubsection{Coalition of sensor groups  $g$  and the query:}

The view of the coalition is defined by
\begin{align}
    V^{\Pi}_{gq} =&  \big(V^{\Pi}_{g},V^{\Pi}_{q}  \big) 
                 = \big( V^{\Pi}_{g},\enc{\hat{c}_{a,k}},\enc{\hat{b}_{a,k}},\hat{c}_{a,k},\hat{G}_{a,k},\hat{A}_{a,k}, \hat{b}_{a,k}, q, F,\nonumber \\
    & Q_k,  \text{coins}_q,\textit{pk}, \text{sk}, \enc{\Gamma_q} \big)\nonumber\\
                 \stackrel{\eqref{eq:vsengrp}}{=}&
                 \big(H_{r,k},R_{r,k}, H_{g,k},y_{g,k},R_{g,k},\bar{c}_{g,k},\bar{G}_{g,k}, \enc{\bar{c}_{g,k}}, \bar{A}_{g,k}, \bar{b}_{g,k},  \enc{\bar{b}_{g,k}},\nonumber \\
    &\enc{\hat{c}_{a,k}},\enc{\hat{b}_{a,k}},\hat{c}_{a,k},\hat{G}_{a,k},\hat{A}_{a,k}, \hat{b}_{a,k}, q, F,   Q_k,\text{coins}_{gq},\textit{pk}, \text{sk},\nonumber \\
    & \enc{\Gamma_{gq}} \big).   
\end{align}

Constructing the simulator $S_{gq}$ from the inputs and outputs of the coalition is done as before
\begin{align}
S_{gq} = \big(&H_{r,k},R_{r,k}, H_{g,k},y_{g,k},R_{g,k},\bar{c}_{g,k},\bar{G}_{g,k}, \widetilde{\enc{\bar{c}_{g,k}}}, \bar{A}_{g,k}, \bar{b}_{g,k},  \widetilde{\enc{\bar{b}_{g,k}}},\nonumber \\
    &\widetilde{\enc{\hat{c}_{a,k}}},\widetilde{\enc{\hat{b}_{a,k}}},\widetilde{\hat{c}_{a,k}}, \hat{G}_{a,k},\hat{A}_{a,k}, \hat{b}_{a,k}, q, F,   Q_k,\text{coins}_{gq},\textit{pk}, \text{sk},\nonumber \\
    & \widetilde{\enc{\Gamma_{gq}}} \big),
\end{align}
which in turn implies that $S_{gq} \stackrel{c}{\equiv} V^{\Pi}_{gq}$. The target of this coalition is to get the constrained zonotopes of the remaining groups. As shown in \eqref{eq:cGgravpr4}, any center and generator of the coalition can determine the containing zonotope of the constrained zonotope. However, the remaining rows of the $\enc{\grave{b}_{a,k}}$ in \eqref{eq:Abgravpr4}, which belongs to the non-colluding sensor group, can not be inferred from the coalition.

\bibliography{ref}


\end{document}